\documentclass[compsoc,conference,a4paper,10pt,times]{IEEEtran}
\usepackage{amsmath,amssymb,amsfonts}
\def\BibTeX{{\rm B\kern-.05em{\sc i\kern-.025em b}\kern-.08em
    T\kern-.1667em\lower.7ex\hbox{E}\kern-.125emX}}

\usepackage{tikz}
\usepackage[T1]{fontenc}  
\usepackage{AlegreyaSans}  
\usepackage{bbold}   
\definecolor{mypurple}{RGB}{150, 10, 75}
\definecolor{mygreen}{RGB}{10, 150, 50}
\definecolor{myblue}{RGB}{10, 10, 200}

\usepackage{enumitem}  
\usepackage{subfigure}
\usepackage{float}
\usepackage{graphicx}
\usepackage{color}
\usepackage{algorithm2e}
\usepackage{hyperref}
\hypersetup{colorlinks, linkcolor=mypurple, urlcolor=myblue, citecolor=mygreen} 

\usepackage{amsthm}   
\usepackage{mathtools} 
\usepackage{thmtools}  
\usepackage{booktabs}    
\usepackage{multirow}
\usepackage[nice]{nicefrac}

\graphicspath{{./figures/}}

\newcommand{\E}{\mathbf{E}}

\newcommand{\argmin}{\operatornamewithlimits{argmin}}

\usepackage{breqn}

\newcommand{\vectr}[1]{\boldsymbol{#1}}

\newcommand{\calx}{\mathcal{X}}   
\newcommand{\calz}{\mathcal{Z}}   
\newcommand{\cala}{\mathcal{A}}   
\newcommand{\call}{\mathcal{L}}   

\newcommand{\calg}{\mathcal{G}}   
\newcommand{\dist}{\mathbf{d}} 

\newcommand{\vtheta}{\vectr{\theta}}     
\newcommand{\vphi}{\vectr{\phi}}           
\newcommand{\vpi}{\vectr{\pi}}           
\newcommand{\vq}{\vectr{q}}           
\newcommand{\vs}{\vectr{s}}           

\newcommand{\leps}{\epsilon}    
\newcommand{\geps}{\varepsilon}  

\newcommand{\vm}{\vectr{m}}     
\newcommand{\vzeros}{\vectr{0}}        
\newcommand{\vones}{\vectr{1}}        

\newcommand{\e} {e}   
\newcommand{\rappor} {RAP}                                          
\newcommand{\inv} {INV}                                          
\newcommand{\ibu} {IBU}

\newcommand{\combr}[1]{[{#1}]\textsubscript{R}} 
\newcommand{\crinv} { \combr{\inv{}} }                                     
\newcommand{\cribu} { \combr{\ibu{}}}                                     
\newcommand{\crrappor} { \combr{\rappor{}}  }                                  

\newcommand{\combm}[1]{[{#1}]\textsubscript{M}} 
\newcommand{\cminv} { \combm{\inv{}} }                                     
\newcommand{\cmibu} { \combm{\ibu{}}}                                     
\newcommand{\cmrappor} { \combm{\rappor{}}  }

\newcommand{\gibu} {GIBU}

\newcounter{ncomm}

\begin{document}

\title{
\vspace{-1cm}
Reconstruction of the distribution of sensitive data under free-will privacy
}
\author{\IEEEauthorblockN{1\textsuperscript{st} Ehab ElSalamouny}
\IEEEauthorblockA{
                               Suez Canal University, Egypt}
\and
\IEEEauthorblockN{2\textsuperscript{nd} Catuscia Palamidessi}
\IEEEauthorblockA{Inria and LIX, \'{E}cole Polytechnique, France}
}
\maketitle
\thispagestyle{plain}  
\pagestyle{plain}  
\begin{abstract}
The local privacy mechanisms, such as k-RR, RAPPOR, and the geo-indistinguishability ones, have become quite popular thanks to the fact that the obfuscation can be effectuated at the users end, thus avoiding the need of a trusted third party. Another important advantage is that each data point is sanitized independently from the others, and therefore different users  may use different levels of obfuscation depending on their  privacy requirements, or they may even use entirely different mechanisms depending on the services they are trading their data for. A challenging requirement in this setting is to construct the original distribution on the users sensitive data from their noisy versions. Existing techniques can only estimate that distribution separately on each obfuscation schema and corresponding noisy data subset. But the smaller are the subsets, the more imprecise the estimations are.
In this paper we study how to avoid the subsets-fractioning problem when combining local privacy mechanisms, thus recovering an optimal utility.  We focus on the estimation of the original distribution, and on the two main methods to estimate it: the matrix-inversion method and the iterative Bayes update. We consider various cases of combination of local privacy mechanisms, and compare the flexibility and the performance of the two methods.
\end{abstract}


\section{Introduction}

Over recent years, there is a growing demand for analyzing amounts of data that are collected from a large number of users. 
To allow this analysis, the users need to release their data which may be sensitive and therefore put their privacy at risk.  
The local privacy model has been presented in the literature to solve this problem \cite{Agrawal:01:PDS,Agrawal:05:ICMD,Duchi:13:Xiv,Kairouz:16:ICML}. 
More precisely, every user applies a \emph{privacy mechanism} that obfuscates his original datum to produce a noisy version of it and then 
sends the latter to the data collector (instead of the original datum). The problem now is estimate statistical properties of the original users' data 
from their noisy releases. In particular, we assume that there is a probability distribution on the values of a sensitive attribute and we want to 
estimate this distribution from the users' noisy releases of this attribute. Existing methods to solve this problem assume that all users apply 
the same privacy mechanism to sanitize their original data. Based on this assumption, the authors of \cite{Duchi:13:Xiv,Kairouz:16:ICML} proposed the 
\emph{matrix inversion} (\inv{}) method in conjunction with the $k$-RR mechanism. A more sophisticated method is the \emph{Iterative Bayesian Update} (\ibu{}) 
which was proposed in \cite{Agrawal:01:PDS,Agrawal:05:ICMD}. This method iteratively computes the maximum likelihood estimate (MLE) for 
the required probability distribution over the alphabet of the sensitive attribute. 

While the above methods work in the setting that all users apply the same privacy mechanism to their original data, we aim in this paper 
to generalize this setting and assume instead that every user applies his arbitrary mechanism. Thus we consider mixtures of different mechanisms that may 
be used. In particular we consider $k$-RR \cite{Kairouz:16:JMLR} and \textsc{Rappor} \cite{Erlingsson:14:CCS} which satisfy local differential privacy.  
In addition we consider the geometric mechanisms \cite{Ghosh:09:STOC,Chatzikokolakis:17:POPETS} which satisfy geo-indistinguishability.    
We also consider Shokri's mechanism \cite{Shokri:12:CCS} that satisfies a particular notion of location privacy. In order to estimate the original distribution we 
consider \gibu{} that was introduced in \cite{Elsalamouny:2020:eurosp} for handling mixtures of mechanisms. We show that this method is 
highly inefficient in its original form, and therefore we provide a more efficient algorithm that produces the same result of \gibu{}. In addition 
we consider various ways of \emph{reusing} \inv{} and \ibu{} in our general setting that consists of different mechanisms. 

We provide an experimental comparisons between the above estimation methods using both synthetic and real data. The results of this comparisons 
are summarized in \autoref{tab:performances1} which shows that estimation accuracy of \gibu{} is superior relative to the other methods.  

\begin{table*}[h]
  \begin{center}
  \caption{performances of construction methods}
  \label{tab:performances1} 
    \bgroup
    \def\arraystretch{1.1}   
  \begin{tabular}{ c | c c c c c} 
  \toprule 
&\crinv{}                                        & \cribu{}                        & \cminv{}     &\cmibu{}      & \gibu{} \\
\midrule                                        
 $k$-RR + $k$-RR      &Bad (Fig.~\ref{fig:cr_KRR_KRR})   & Bad (Fig.~\ref{fig:cr_KRR_KRR})  & Excellent (Fig.~\ref{fig:cm_KRR_KRR_INV})  & Good (Fig.~\ref{fig:cm_KRR_KRR_IBU}) &Excellent \\
 Geometric + Geometric  &Bad (Fig.~\ref{fig:cr_GEOM_GEOM})      &Bad (Fig.~\ref{fig:cr_GEOM_GEOM})      & Bad (Fig.~\ref{fig:cm_GEOM_GEOM_INV}) &Bad (Fig.~\ref{fig:cm_GEOM_GEOM_IBU}) &Excellent \\
 $k$-RR + Geometric &Bad (Fig.~\ref{fig:cr_VLGEOM_KRR}) &Bad (Fig.~\ref{fig:cr_VLGEOM_KRR}) &Bad (Fig.~\ref{fig:cm_GEOM_KRR_INV}) &Bad (Fig.~\ref{fig:cm_GEOM_KRR_IBU}) &Excellent \\
 Shokri +Shokri         &NA                    &Bad (Fig.~\ref{fig:cr_SH_SH})    &NA   &Bad (Fig.~\ref{fig:cm_SH_IBU}) &Excellent \\
  \bottomrule 
  \end{tabular}
  \egroup
  \end{center}
\end{table*}

\begin{table}[h]
  \begin{center}
  \caption{construction methods under \textsc{Rappor} mechanisms}
  \label{tab:performances2} 
    \bgroup
    \def\arraystretch{1.1}   
  \begin{tabular}{ c | c c c } 
  \toprule 
                                  &\crrappor{}                                        & \cmrappor{}         & \gibu{} \\
\midrule 
High privacy  & Bad (Fig.~\ref{fig:cr_RAP_P0.1_RAP_P1.0}) &Excellent (Fig.~\ref{fig:cm_RAP_P0.1_RAP_P1.0}) &Excellent \\
Low privacy  &Bad (Fig.~\ref{fig:cr_RAP_P1.0_RAP_P10.0})  &Good (Fig.~\ref{fig:cm_RAP_P1.0_RAP_P10.0})  &Excellent   \\                         
  \bottomrule 
  \end{tabular}
  \egroup
  \end{center}
\end{table}

\begin{table}[h]
\begin{center}  
\caption{Notations}
\label{tab:notations}
\begin{tabular}{l l}  
\toprule
$\calx$ & an alphabet of secrets of users. \\ 
$\vtheta$ & original probability distribution on $\calx$. \\
$[n]$ & the set $\{1,2,\dots,n\}$ referring to users.\\
$X^i$ & random var describing the secret of user $i$. \\
$A^i$ & the privacy mechanism applied by user $i$. \\
$Z^i$ & random var describing the (noisy) observable of $A^i$. \\
$\calz^i$ & the alphabet of $Z^i$, i.e. possible observables of $A^i$. \\
$\vm^i$ & marginal distribution of $Z^i$ (over the alphabet $\calz^i$). \\
$A[n]$  & the average of mechanisms $A^i$ for $i\in[n]$.\\ 
$\vq[n]$ & empirical distribution for $n$-size noisy data. \\
$\hat\vtheta[n]$ & estimated value for the original distribution $\vtheta$. \\

\midrule

\rappor{} & the estimator under a \textsc{Rappor} mechanism\\
\inv{} & matrix inversion estimator \\
\ibu{} & iterative Bayesian update\\

\midrule

\combr{\e{}} & combining the results of estimator \e{}\\ 
%
\combm{\e{}} & applying estimator \e{} to a compound mechanism.\\ 
%
\gibu{} & generalized iterative Bayesian update \\
\bottomrule
\end{tabular}
\end{center}
\end{table}

\subsection{Contributions}
\begin{itemize}

\item{Starting with the situation that the mechanisms of the users are various, 
but restricted to have the same signature, 
we extend the classical matrix inversion method \cite{Agrawal:05:ICMD}, 
known also as the empirical estimator \cite{Kairouz:16:ICML}, to work in this setting instead of operating 
under the assumption that all mechanisms are identical. We call this method \cminv{}.}

\item{When all users are restricted to apply only $k$-RR mechanisms, but with arbitrary levels of privacy, we prove that 
\cminv{} is consistent in the sense that its estimated distribution converges \emph{in probability} to the real distribution. 
Furthermore, we derive an upper bound on the $\ell_2^2$ estimation error.}

\item{Similarly we extend the local privacy model under Google's \textsc{Rappor} mechanisms \cite{Erlingsson:14:CCS} to allow every 
user to set his own privacy level. Given this situation we extend the standard estimation procedure under 
\textsc{Rappor}. We also derive an upper bound on its $\ell_2^2$ error showing that the new estimator is also consistent.}

\item{Abstracting away from the above restrictions, we consider the most general model of local privacy in which the users 
apply their own mechanisms, that may vary both in signatures and in privacy guarantees. In this `privacy-liberal' scenario we 
provide a compositional and scalable algorithm \gibu{} that accommodates all these various mechanisms in the estimation process.}

\item{We experimentally show that the estimation performance of \gibu{} is better compared to \cminv{}.}


\item{Since \cminv{} usually requires a post-processing step to obtain a valid distribution, we describe an additional method 
\cmibu{} that uses the average mechanism to yield an estimated distribution. However we show that this method in some cases, e.g. with Shokri's mechanism \cite{Shokri:12:CCS}, may not converge 
(with large number of samples) to the real distribution.} 

\item{We compare \cmibu{} to \gibu{} using different mechanisms, e.g. Geometric, $k$-RR, and Shokri's showing that \gibu{} 
is consistently superior.}

\end{itemize}
\section{Preliminaries}

\subsection{$k$-ary randomized response mechanisms}\label{sec:krr}

The $k$-ary randomized response mechanism, abbreviated as $k$-RR, obfuscates every datum from the alphabet 
$\calx$, with $|\calx| = k$, to produce a noisy observable from the same alphabet, i.e. $\calz = \calx$. 
This mechanism was originally introduced by Warner \cite{Warner:65:jastat} for binary alphabets, and was later extended 
by \cite{Kairouz:16:JMLR} to arbitrary $k$-size alphabets. 
Given a privacy parameter $\leps>0$, this mechanism applied to a datum $x$ produces an observable 
$z$ with probability  
\begin{equation}\label{eq:krr}
P(z|x) = \frac{1}{k-1+ e^\leps} 
\left\{ 
\begin{array}{l l}
e^\leps & \mbox{if $z=x$}\\
1                & \mbox{if $z\neq x$} 
\end{array} 
\right.
\end{equation} 
%
It follows from the definition (\ref{eq:krr}) that the $k$-RR mechanism satisfies $\leps$-local differential privacy.

\subsection{Geometric mechanisms}\label{sec:geometric}

Suppose that the alphabet of secrets $\calx$ is a bounded linear range of integers between $r_1$ and $r_2$ inclusive, 
where $r_1<r_2< \infty$. Then the truncated geometric mechanism \cite{Ghosh:09:STOC}, with parameter $\geps$, maps every $x\in\calx$ 
to an integer $z$ in the same alphabet $\calx$ with probability 
\begin{align} \label{eq:tgeometric}
P(z |x) &= c_z \, e^{-\geps |z-x|},\\
\mbox{where}\; \; c_z &= \frac{1}{1+e^{-\geps}} 
\left\{ 
 \begin{array}{l l}
1            &\mbox{if $z \in \{r_1, r_2\}$} \\
1-e^{-\geps} &\mbox{if $r_1< z< r_2$} \\
0            & \mbox{otherwise}
\end{array} \label{eq:tgeom-cz}
\right..
\end{align}
For the planar alphabet in which the data points are the cells of a grid, we will use a \emph{planar} variant of the 
truncated geometric mechanism which we describe in the following subsection. 

\subsection{Planar geometric mechanisms}
\label{sec:plan_geom}
Suppose that the planar space is discretized by an infinite grid of squared cells, where the side length
of each cell is $s$. Let $\calg$ be the set of centers of these cells. Then every 
element of $\calg$ is indexed by its coordinates $(i,j) \in \mathbb{Z}^2$. For any $\calx \subseteq \calg$, 
a \emph{planar geometric mechanism} (parametrized by $\geps,s$) reports a point $z\in\calg$ 
from a real location $x\in\calx$ according 
to the probability 
\begin{align*}
P(z | x) &= \lambda\, e^{-\geps\, \dist( x , z )} \qquad x\in \calx, z \in \calg  \\
	        \text{where }
	        \lambda &= 1 / \sum_{(i,j) \in \mathbb{Z}^2} e^ {- \geps\, s \, \sqrt{i^2+j^2}} 
\end{align*}
and $\dist(\cdot,\cdot)$ is the planar (i.e. Euclidean) distance. 
If $\calx$ is finite we can define a \emph{truncated} version of the above geometric mechanism. Basically it is obtained by drawing  points in $\calg$ 
according to  the above distribution and then remapping each of them to its 
nearest point in $\calx$.  

Since the alphabet $\calx$ in this paper are bounded, we will ignore the term `truncated' when we 
use the geometric mechanisms, and refer to them directly as linear and planar geometric mechanisms. 
We remark that geometric mechanisms are known to satisfy $\geps$-geo-indistinguishability 
\cite{Andres:13:CCS}, in which the distinguishability between two points $x, x'$ in the space of secrets $\calx$ 
is bounded by $\geps \dist(x,x')$, where $\dist(\cdot,\cdot)$ is the Euclidean distance. 
This notion differs from the $\leps$-local differential privacy \cite{Duchi:13:FOCS,Kairouz:16:JMLR} because 
in the latter notion the distinguishability between any two points is bounded by the fixed $\leps$.

\section{Local privacy model}
\begin{figure}[h]
\center
\includegraphics{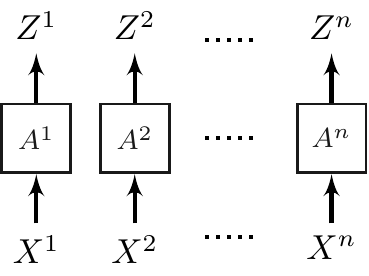}
\caption{The basic local privacy model. Every user datum $X^i$ is obfuscated locally by a mechanism $A^i$
to yield the noisy output $Z^i$.}
\label{fig:local_priv_model}
\end{figure}
We consider $n$ users, each of them is labelled by $i \in [n]$. The sensitive datum of user 
$i$ is described by the random variable $X^i$. We assume that these sensitive data are 
drawn i.i.d from the alphabet of secrets $\calx$ according to a hidden distribution $\vtheta$. 
As shown in Figure \ref{fig:local_priv_model}, every user obfuscates his datum $X^i$ 
using an arbitrary privacy mechanism $A^i$ to yield a noisy observation $Z^i$ taking values from 
an alphabet of observables $\calz^i$ which depends indeed 
on the applied mechanism. 

As a consequence of the above process, each random observation $Z^i$ has a \emph{marginal} 
probability distribution $\vm^i$ that depends on the real distribution $\vtheta$ and the 
obfuscation mechanism $A^i$; namely $\vm^i = \vtheta\, A^i$. In particular $m^i_z$ 
is the probability of observing $z \in \calz^i$ from the user $i$.
Note that the observations $Z^i$ have in general different marginal distributions, 
and possibly different alphabets $\calz^i$ depending on the corresponding mechanisms. 
This makes estimating the original distribution $\vtheta$ from the observations challenging, 
compared to the traditional case, e.g. \cite{Agrawal:01:PDS,Agrawal:05:ICMD,Duchi:13:Xiv,Kairouz:16:ICML}, 
in which all observations are produced by the same mechanism and therefore follow the same 
marginal distribution. 
%

\section{Estimation by combining results}\label{sec:combining_estimates}

Estimation methods existing in the literature of local privacy work under the assumption 
that all the noisy data are produced by the same mechanism. In the following we describe 
one approach of reusing these methods in our case which involves a set $\cala$ of different 
mechanisms applied by the users. 
The available noisy data are regarded as a collection of disjoint subsets $\{D^A: A \in \cala\}$, 
where each subset $D^A$ consists of $n^A$ observations produced by a mechanism $A$. 
Note that $n = \sum_{A\in\cala} n^A$. Then using a generic estimator $\mathbf{e}$, the method
is described by Equations (\ref{eq:generic_e}) and (\ref{eq:combining_results}).
%
\begin{align}
\hat\vtheta[A] &= \mathbf{e}(A, D^A) \quad\forall A\in\cala  \label{eq:generic_e},\\
\hat\vtheta &= \sum_{A\in\cala} \frac{n^A}{n} \hat\vtheta[A]. \label{eq:combining_results}
\end{align}
%
%
The expression $\mathbf{e}(A, D^A)$ in (\ref{eq:generic_e}) is an estimate resulting from 
applying $\mathbf{e}$ to the mechanism $A$ and its generated noisy data $D^A$. 
Then by (\ref{eq:combining_results}) the overall estimate $\hat\vtheta$ is a weighed 
average of the estimates $\hat\vtheta[A]$ for all $A\in\cala$ using the proportions of their 
underlying data. 

While the above method takes advantage of existing `off-the-shelf' estimators developed for local privacy, 
notice that each instance of the underlying estimator $\mathbf{e}$ works only on a subset of noisy data. 
This impacts the estimation performance as shown experimentally in Section \ref{sec:cresults_vs_gibu}. 
\section{Estimation by combining mechanisms}\label{sec:combining_mechanisms}
Instead of combining the estimates from disjoint subsets of the noisy data as in 
Section \ref{sec:combining_estimates}, we develop in this section 
estimators that rely on combining the underlying users' mechanisms into one. 
Then the used estimator is applied on the compound mechanism instead of applying it individually on each underlying mechanism. 

\subsection{Compound-mechanism inversion \cminv{}}\label{sec:cminv}
In the following we assume that the mechanisms of the users have the same alphabet $\calz$ of 
observables. In this case, we can construct the empirical distribution $\vq[n]$ 
where the $z$-th component $q[n]_z$ is the proportion of $z\in\calz$ in the noisy data, that is 
\begin{equation}\label{eq:q}
q[n]_z = \frac{1}{n} \sum_{i\in[n]} \mathbb{1}_{Z^i = z} \quad \forall z\in\calz. 
\end{equation}
%
Then we define the estimate of the \emph{compound-mechanism inversion} estimator \cminv{} as follows. 
\begin{equation}\label{eq:comb_mat_inv}
\hat\vtheta[n] = \vq[n] \, (A[n])^{-1} \quad \textit{where}\quad A[n] = \frac{1}{n}\sum_{i=1}^n A^i \, .  
\end{equation}
As a special case, if all users apply the same mechanism $A$, then clearly $A[n] = A$ 
and (\ref{eq:comb_mat_inv}) coincides, in this case, with the matrix inversion estimator 
(also known as the empirical estimator) \inv{}
which was described by \cite{Agrawal:05:ICMD,Kairouz:16:ICML}. 
%

From the definition (\ref{eq:comb_mat_inv}) of \cminv{} the resulting estimate $\hat\vtheta[n]$ 
may in general contain negative components. 
Therefore we post-process it using two methods to yield a valid (non-negative) distribution. 
The first method is projecting $\hat\vtheta[n]$ onto the probability simplex, 
with respect to the $\ell_2$ distance \cite{Wang:2013ProjectionOT}. The second method is truncating the negative 
components of $\hat\vtheta[n]$ to $0$ and then normalizing the resulting vector. 

A fundamental property that describes the consistency of \cminv{} is shown by the following proposition. 
%
\begin{restatable}{proposition}{cminvconv}
\label{prop:mat_inv_conv}
Suppose that $A^i$, for $i\in[n]$, have the same alphabet of observables and $A[n]$ is invertible for every $n>0$. 
Let $\hat\vtheta[n]$ be the result of \cminv{} (\ref{eq:comb_mat_inv}). Then  
$
(\hat\vtheta[n] - \vtheta)\, A[n] \to \vzeros
$
almost surely. Furthermore if $A[n]$ converges to some $\bar A$, then $\hat\vtheta[n] \to \vtheta$ almost surely. 
\end{restatable}
In particular, if all users apply the same mechanism, or they choose their mechanisms from a finite set according to 
some probability distribution, then $\hat\vtheta[n] \to \vtheta$ almost surely.  
Moreover we will show in Section \ref{sec:inv_krr} that if all mechanisms are $k$-RR, with various levels of privacy, the expected
$\ell_2^2$ error of \cminv{} (\ref{eq:comb_mat_inv}) converges with large $n$ to $0$ and 
therefore $\hat\vtheta[n] \to \vtheta$ \emph{in probability}.  

\subsection{\cminv{} under various $k$-RR mechanisms}\label{sec:inv_krr}
Suppose that every user $i\in [n]$ sanitizes his datum using a $k$-RR (\ref{eq:krr}), 
with a privacy level $\leps_i$ of his choice. In this scenario we can express the estimate of \cminv{} as follows. 
First we construct the average mechanism $A[n]$ using (\ref{eq:krr}). Note in this case that $A[n]$ is also a $k$-RR 
mechanism with $\leps[n]$ that satisfies 
\begin{equation}\label{eq:avg_krr}
\frac{1}{k-1+ e^{\leps[n]}} = \frac{1}{n}\sum_{i=1}^n \frac{1}{k-1 + e^{\leps_i}}.
\end{equation}
Then by (\ref{eq:comb_mat_inv}) the estimate $\hat\vtheta[n]$ of \cminv{} is 
obtained by solving the equation $\hat\vtheta[n] \, A[n] = \vq[n]$, which yields  
\begin{equation}\label{eq:comb_mat_inv_krr}
\hat\vtheta[n] = \frac{e^{\leps[n]} + k - 1}{e^{\leps[n]} -1} \, \vq[n] - \frac{1}{e^{\leps[n]}-1}\, \vones.
\end{equation}
Using the above expression of the estimate $\hat\vtheta[n]$ we can derive an upper bound on the 
expected $\ell_2^2$ error of \cminv{} under various $k$-RR mechanisms as follows. 
\begin{restatable}{proposition}{krrmatinvloss}
\label{prop:krr_mat_inv_loss}
Suppose that every user $i \in [n]$ applies a $k$-RR mechanism with 
arbitrary $\epsilon_i>0$. Let $\leps[n]$ be defined by (\ref{eq:avg_krr}). 
Then the estimate $\hat\vtheta[n]$ of \cminv{} satisfies 
\begin{align*}
\E[ {\| \hat\vtheta[n] - \vtheta \|}_2^2 ] &\leq \frac{1 - \sum_{x\in\calx} {\theta_x}^2}{n} \\
                    &\qquad+ \frac{k -1}{n} \left( \frac{k+2(e^{\leps[n]} -1)}{(e^{\leps[n]} - 1)^2}\right)
\end{align*}
where the equality holds if the values $\leps_i$, for all $i\in [n]$, are equal.    
\end{restatable}
The above proposition is general in the sense that it allows the privacy level $\leps_i$ for 
every user to be arbitrary according to his preference. As a special case when $\leps_i$ for all $i\in[n]$ 
are identical, i.e. all users apply the same $k$-RR mechanism, \autoref{prop:krr_mat_inv_loss} yields 
the same error that was reported by \cite{Kairouz:16:ICML} under this assumption. 

Another important consequence of \autoref{prop:krr_mat_inv_loss} is that 
if $\inf_n \leps[n] >0$, the estimation error given by \autoref{prop:krr_mat_inv_loss} converges 
to $0$, which implies by Markov's inequality that the \cminv{} estimator under $k$-RR mechanisms 
(\ref{eq:comb_mat_inv_krr}) is consistent, i.e. $\hat\vtheta[n] \to \vtheta$ in probability. 
\footnote{
This means that 
$
\lim_{n\to\infty} P({\| \hat\vtheta[n] - \vtheta \|}_2 > \delta) = 0 
$ 
for all $\delta>0$. In fact the condition $\inf_n \leps[n] > 0$ 
implies that the expected error in 
\autoref{prop:krr_mat_inv_loss} converges (as $n$ grows) to $0$, which implies the 
above equation by Markov's inequality.
}

Finally we can see that the estimate of \cminv{} under $k$-RR mechanisms and $e^{\leps[n]} \geq k$ 
converges in the same order as the best estimator. 
In fact, a lower bound for the $\ell_2^2$ estimation error of any estimator is $(1- \sum_{x\in \calx} \theta_x^2)/n$  
which corresponds to the maximum likelihood estimator working on non-sanitized samples \cite{Kamath:15:pmlr}. 
From \autoref{prop:krr_mat_inv_loss}, and assuming that $e^{\epsilon[n]} \geq k$, the error of \cminv{} 
is above this bound by at most $(3+1/(k-1))/n$. 

%
%


\subsection{Compound-matrix iterative Bayesian update \cmibu{}}\label{sec:cmibu}

Assuming that all users apply the same mechanism $A$, 
the authors of \cite{Agrawal:01:PDS} proposed a procedure called iterative Bayesian update (\ibu{}) 
to estimate the original distribution $\vtheta$. This procedure starts 
with a full-support distribution $\vtheta^0$ on $\calx$ and then using the empirical distribution 
$\vq[n]$ (on $\calz$) defined by (\ref{eq:q}), the estimate is refined iteratively as follows
\begin{equation} 
\theta^{t+1}_x = \sum_{z\in\calz} q[n]_z\, \frac{\theta^t_x A_{xz}}{\sum_{u\in\calx} \theta^t_u A_{uz}} \quad \forall x\in\calx.
\end{equation}
%
%
The procedure terminates and return $\vtheta^t$ when two successive estimates are close enough. \ibu{} coicides
with \gibu{} (cfr.~Section \ref{sec:gibu}) when the users apply the same mechanism $A$, and in this case \ibu{} 
returns the MLE for $\vtheta$.  

In our scenario where various mechanisms are used (instead of a fixed one),  one way of reusing \ibu{} is to construct the average 
mechanism $A[n]$ defined by (\ref{eq:comb_mat_inv}) and then apply \ibu{} to $A[n]$ and $\vq[n]$. We will call this method 
\cmibu{}. Note that this method (like \cminv{}) requires that all the mechanisms have the same alphabet of observables $\calz$
to allow computing $A[n]$. 

Compared to \cminv{}, one advantage of \cmibu{} is that $A[n]$ is not needed to be invertible.
In addition, \cmibu{} does not require post-processing because it always returns a valid distribution. 
However the main limitation of \cmibu{} is that it has no formal guarantees. In particular it does not yield the MLE 
unless all mechanisms are identical. 

More seriously, this method may converge \emph{badly} or may not converge at all when the 
average mechanism $A[n]$ turns to be non-informative. Consider for example the following two 
mechanisms
\begin{equation}\label{eq:non-informative}
A =\begin{bmatrix}
    \nicefrac{3}{4} &     \nicefrac{1}{4}   \\
    \nicefrac{1}{4} &     \nicefrac{3}{4} 
\end{bmatrix}, 
\qquad
A' =\begin{bmatrix}
    \nicefrac{1}{4} &     \nicefrac{3}{4}   \\
    \nicefrac{3}{4} &     \nicefrac{1}{4} 
\end{bmatrix}, 
\end{equation} 
It is clear that both $A$ and $A'$ provide a reasonable statistical utility if only one of them is applied by all users. 
However, if half of the users apply $A$ and the other half apply $A'$, then the average mechanism $A[n]$ 
is entirely non-informative because each row turns to be a uniform distribution over the observables, 
This of course makes estimating the original distribution using \cmibu{} extremely inaccurate in this particular case.

\subsection{Compound-\textsc{Rappor} estimator \cmrappor{}}\label{sec:cm_rappor_estimator}


The mechanism \textsc{Rappor} is built on the idea of randomized response to allow collecting statistics from 
end-users with differential privacy guarantees \cite{Erlingsson:14:CCS}. 
The simplest version of this mechanism, known as Basic One-Time \textsc{Rappor}, obfuscates every user's private 
datum $X^i$ (taking a value from the alphabet $\calx$) as follows. $X^i$ is encoded in a `one-hot' binary vector 
$B^i \in \{0,1\}^{|\calx|}$ such that $B^i_x = 1$ if $X^i=x$ and $B^i_x = 0$ otherwise. Then given a privacy parameter $\leps>0$, every bit $B^i_x$ is obfuscated independently to a random bit $V^i_x$ as follows.
\begin{equation}\label{eq:rappor}
V^i_x =  
\left\{ 
\begin{array}{l l}
B^i_x & \mbox{with probability $\frac{e^{\leps/2}}{1+e^{\leps/2}}$ ,}\\
1 - B^i_x   & \mbox{with probability $\frac{1}{1+e^{\leps/2}}$ .}
\end{array} 
\right.
\end{equation} 
Finally, the random binary vector $V^i\in \{0,1\}^{|\calx|}$ resulting from the above obfuscation scheme is reported 
to the server. As shown by \cite{Duchi:13:FOCS,Kairouz:16:JMLR}, the \textsc{Rappor} mechanism satisfies 
$\leps$-local differential privacy. 


It is clear that the observables of \textsc{Rappor} are 
bit-vectors drawn from $\calz = \{0,1\}^{|\calx|}$ which may be too large (of size $2^{|\calx|}$), 
making it impractical to obtain the inverse matrix required to apply the \inv{} estimator. 
The authors of \cite{Kairouz:16:ICML} described a special estimator under \textsc{Rappor}, but 
valid only when all users set the privacy parameter $\leps$ to the same value. In the following we 
remove this restriction and present the combound-\textsc{Rappor} estimator which we coin as \cmrappor{}. 

Suppose that each user $i\in[n]$ applies \textsc{Rappor} with an arbitrary $\leps_i$. Then we define $\leps[n]$ to satisfy
\begin{equation}\label{eq:avg_rappor}
\frac{1}{1+ e^{\nicefrac{\leps[n]}{2}}} = \frac{1}{n}\sum_{i=1}^n \frac{1}{1 + e^{\nicefrac{\leps_i}{2}}}.
\end{equation}
Using the vectors $V^i$ reported from individual users, we define 
the vector $\vs[n] = \nicefrac{1}{n} \sum_{i=1}^n V^i$. Note that $\vs[n]$ has length 
equal to the size $\calx$ (the alphabet of secrets), and $s[n]_x$ is the proportion of 
vectors $V^i$ having the $x$-th bit set to $1$. 
Finally we define the estimate of \cmrappor{} as 
\begin{equation}\label{eq:rappor_estimator}
\hat\vtheta[n] = \frac{e^{\nicefrac{\leps[n]}{2}} + 1}{e^{\nicefrac{\leps[n]}{2}} -1} \, \vs[n] - \frac{1}{e^{\nicefrac{\leps[n]}{2}}-1}\, \vones.
\end{equation}
Since the resulting estimate $\hat\vtheta[n]$ may contain negative components, it is 
post-processed by projection or normalization (similar to \cminv{} in Section \ref{sec:cminv})
to obtain a valid distribution.   
A special case of \cmrappor{} is obtained when all the mechanisms are identical with parameter $\leps$. 
In this case, by (\ref{eq:avg_rappor}), we have $\leps[n] = \leps$, 
and (\ref{eq:rappor_estimator}) coincides with the basic estimator \rappor{} proposed by 
\cite{Duchi:13:Xiv,Erlingsson:14:CCS}.    
 
The quality of \cmrappor{} can be described analytically as the expected $\ell_2^2$
between the real and estimated distributions as follows. 
\begin{restatable}{proposition}{rapporestimatorloss}
\label{prop:rappor_estimator_loss}
Suppose that every user $i \in [n]$ applies a \textsc{Rappor} mechanism with 
arbitrary $\leps_i>0$. Let $\leps[n]$ be defined by (\ref{eq:avg_rappor}). 
Then the estimate $\hat\vtheta[n]$ of \cmrappor{} satisfies 
%
\[
\E[ {\| \hat\vtheta[n] - \vtheta \|}_2^2 ] \leq \frac{1 - \sum_{x\in\calx} {\theta_x}^2}{n}
                    + \frac{1}{n} \frac{k e^{\nicefrac{\leps[n]}{2}} }{(e^{\nicefrac{\leps[n]}{2}} - 1)^2}
\]
where the equality holds if the values $\leps_i$, for all $i\in [n]$, are equal.    
\end{restatable}
It follows from the above result that if $\inf_n \leps[n] >0$, the estimation error of \cmrappor{} 
converges to $0$, which implies by Markov's inequality that this estimator is consistent in the sense 
that $\hat\vtheta[n] \to \vtheta$ in probability.
\section{Generalized iterative Bayesian update}\label{sec:gibu}
A rigorous approach to reconstruct the original distribution $\vtheta$ on $\calx$ is to 
compute the maximum likelihood estimate (MLE) of this distribution from the observed noisy data. 
The authors of \cite{Elsalamouny:2020:eurosp} developed 
an iterative algorithm, called \gibu{}, that evaluates this MLE. 
Basically, in each iteration, the current estimate $\vtheta^t$ for the distribution over $\calx$ 
is refined to $\vtheta^{t+1}$ as follows.
\begin{equation}\label{eq:gibu}
\theta^{t+1}_x = \frac{1}{n}\sum_{i=1}^n   \frac {\theta^t_x G_{x i}}{\sum_{u \in \calx} \theta^t_u G_{u i}} \quad \forall x\in\calx
\end{equation}
where $G_{x i}$ is the conditional probability of the observation from user $i$ given that his real private datum 
is $x\in \calx$. 
This procedure is highly inefficient since each iteration requires time of order 
$n |\calx|^2$, and $n$ (the number of users) may be arbitrarily large. 
Therefore we propose a refined algorithm which is scalable. 
Let $\cala$ be the set of different mechanisms applied by the users. Then for every 
mechanism $A\in \cala$ we have the following triplet:
\begin{itemize}
\item the number of users $n^A$ who used the mechanism $A$,
\item the alphabet of observables $\calz^A$ for $A$, 
\item the empirical distribution over $\calz^A$, written as $\vq^A$. 
\end{itemize}
Note that the total number of users is $n = \sum_{A\in\cala} n^A$. Using the above triplet 
for every mechanism in $\cala$, \autoref{alg:gibu} describes \gibu{} which returns an MLE for $\vtheta$. 

\begin{algorithm}
\SetKwInOut{Input}{inputs}
\Input{
$\cala$ \tcp*{the set of mechanisms}\\  
$(\calz^A, \vq^A, n^A)\,\forall A \in \cala$, and $\delta>0$
}
$n \gets \sum_{A\in\cala} n^A$ \;
\tcp{normalized log-likelihood}
$\call(\vphi) = \sum_{A\in \cala}  \frac{n^A}{n} \sum_{z\in\calz^A} q_z^A\log \sum_{x\in\calx} \phi_x A_{xz}$ \;
$t \gets 0$\; 
$\vtheta^0 \gets \textit{any distribution on $\calx$ s.t.}\,\, \theta^0_x>0\, \forall x \in \calx$\; 
\Repeat{ $| \call(\vtheta^t)  - \call(\vtheta^{t-1}) | < \delta$ }{
$\forall x\in\calx:$ \\ 
$\quad \theta^{t+1}_x \gets \sum_{A\in\cala}\frac{n^A}{n} \sum_{z\in\calz^A} q_z^A \frac{\theta^t_x A_{xz}}{\sum_{u\in\calx} \theta^t_u A_{uz}}$\; 
$t \gets t+1$ \;
}
\Return $\vtheta^t$
\vspace{3mm}
\caption{\gibu{} working on a set $\cala$ of mechanisms.}
\label{alg:gibu}
\end{algorithm}

\begin{restatable}{theorem}{gibumle}
\label{thm:gibu_mle}
\autoref{alg:gibu} returns an MLE for $\vtheta$. 
\end{restatable}

It can be seen that each iteration of \autoref{alg:gibu} requires time in order of 
$\sum_{A\in\cala} |\calz^A| \cdot |\calx|^2$ which is independent of $n$ and therefore 
shows a significant reduction in the time complexity compared to (\ref{eq:gibu}). 
We also note that every iteration is a sum of terms that can be computed in parallel since 
each term depends only on the triplet $(\calz^A, \vq^A, n^A)$ of the mechanism $A$. 
In conclusion, \autoref{alg:gibu} is scalable, allowing us to run it 
over large samples and compare it to other methods as we proceed in the following sections. 

%
%

\section{Evaluation of combining results}\label{sec:cresults_vs_gibu}

In this section we experimentally evaluate the approach of combining results described in Section \ref{sec:combining_estimates}.
We consider three off-the-shelf estimators that may replace $\mathbf{e}$ in (\ref{eq:generic_e}), 
namely the standard estimator under the \textsc{Rappor} mechanism (\rappor{}) \cite{Duchi:13:Xiv,Erlingsson:14:CCS}, 
\inv{} \cite{Agrawal:05:ICMD,Kairouz:16:ICML}, and \ibu{} \cite{Agrawal:01:PDS}. We recall that each one of these methods
works on noisy data produced by one mechanism as described respectively in Sections \ref{sec:cm_rappor_estimator}, 
\ref{sec:cminv}, and \ref{sec:cmibu}. 

We consider an alphabet $\calx =  \{0,1,\dots, |\calx|-1 \}$, and then we construct the real data of the users synthetically 
by sampling from a binomial distribution on $\calx$ with $\alpha = 0.5$. 

\begin{figure}[h]
\centering 
\subfigure[high privacy regime: $\leps_i$ between $0.1$ and $1.0$]{
      \label{fig:cr_RAP_P0.1_RAP_P1.0}
      \includegraphics[width=0.5\textwidth]{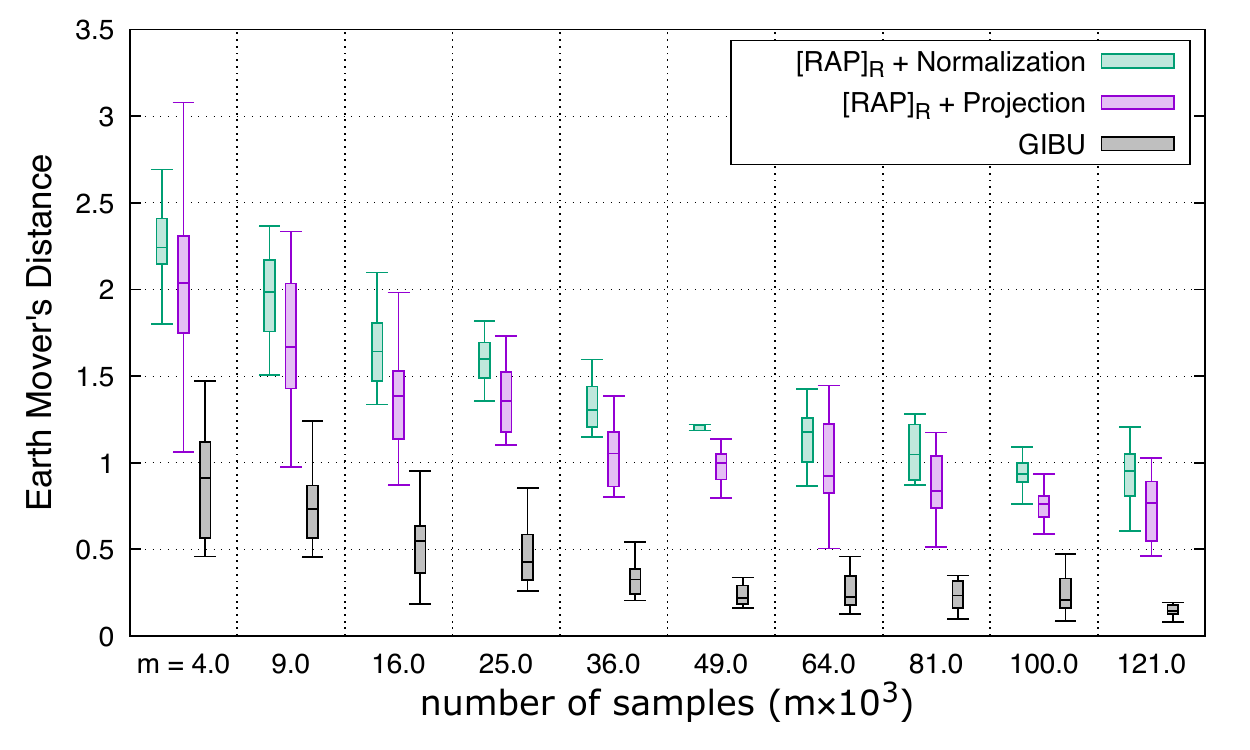}
      }
\subfigure[low privacy regime: $\leps_i$ between $1.0$ and $10.0$]{
      \label{fig:cr_RAP_P1.0_RAP_P10.0}
      \includegraphics[width=0.5\textwidth]{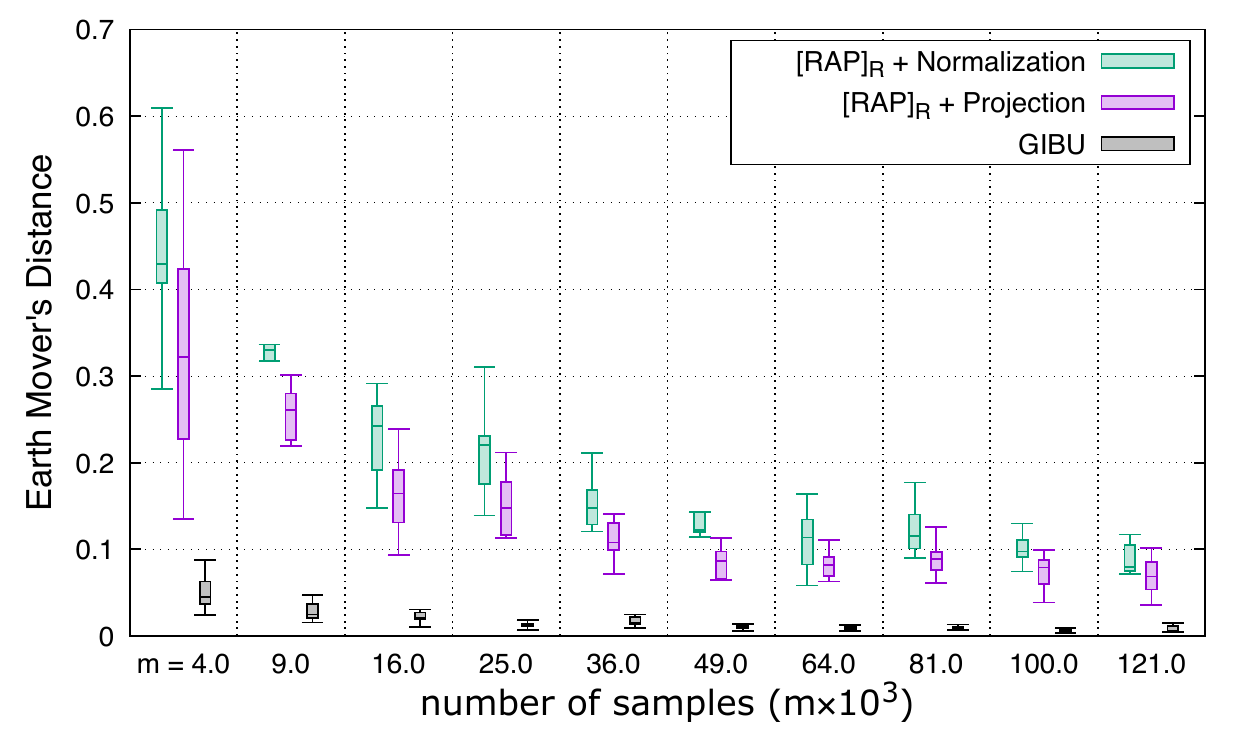}
      }
\caption{Performance of \crrappor{} and \gibu{} for $\calx = \{0,1,\dots,19 \}$.
The noisy data are produced by 10 \textsc{Rappor} mechanisms having various $\leps_i$ 
in high privacy regime (a), and in low privacy regime (b) (cfr.~\autoref{tab:rappor_rappor}).}
\label{fig:cr_RAPPOR_RAPPOR}
\end{figure}

In the experiments of this section we will use \textsc{Rappor} mechanisms (\ref{eq:rappor}) and $k$-RR mechanisms (\ref{eq:krr}). 
We also use the truncated linear geometric mechanism \cite{Ghosh:09:STOC} described in Section \ref{sec:geometric}. 
We start our experiments by letting the original data be obfuscated by $10$ \textsc{Rappor} mechanisms with various $\leps_i$, 
and then we apply Equations (\ref{eq:generic_e}) and (\ref{eq:combining_results}) with $\mathbf{e}$ 
replaced by \rappor{}. Hence we call the method in this case \crrappor{}. Recall that every result of \rappor{} 
requires post-processing by \emph{projection} or \emph{normalization} to return a valid distribution (cfr.~Section \ref{sec:cm_rappor_estimator}).  Figure \ref{fig:cr_RAPPOR_RAPPOR} shows the estimation performance 
of \crrappor{} with these two post-processing methods, and also that of \gibu{}. The performance is measured 
by the earth mover's distance (EMD) between the original and estimated distributions.
\footnote{
The earth mover's distance between two probability distributions $\theta, \phi$, on a set $\calx$, is 
defined as the minimum cost of transforming $\theta$ into $\phi$ 
by transporting the masses of $\theta$ between the elements of $\calx$. 
}
 
%

\begin{figure}[h]
\centering 
\includegraphics[width=0.50\textwidth]{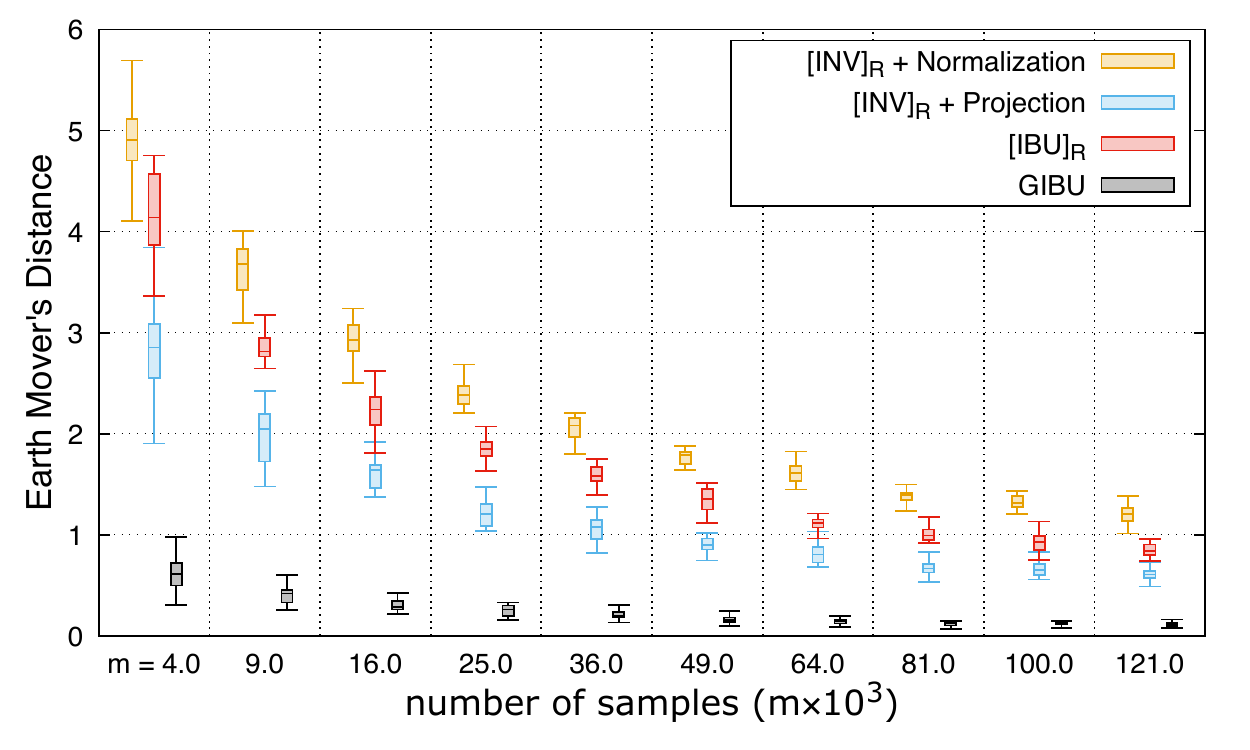}
\caption{Performance of \crinv{}, \cribu{} and \gibu{} for a linear alphabet of secrets $\{0,1,\dots,99 \}$
The noisy data are produced using 10 different $k$-RR mechanisms (cfr.~\autoref{tab:krr_krr}).}
\label{fig:cr_KRR_KRR}
\end{figure}

\begin{figure}[h]
\centering 
\includegraphics[width=0.50\textwidth]{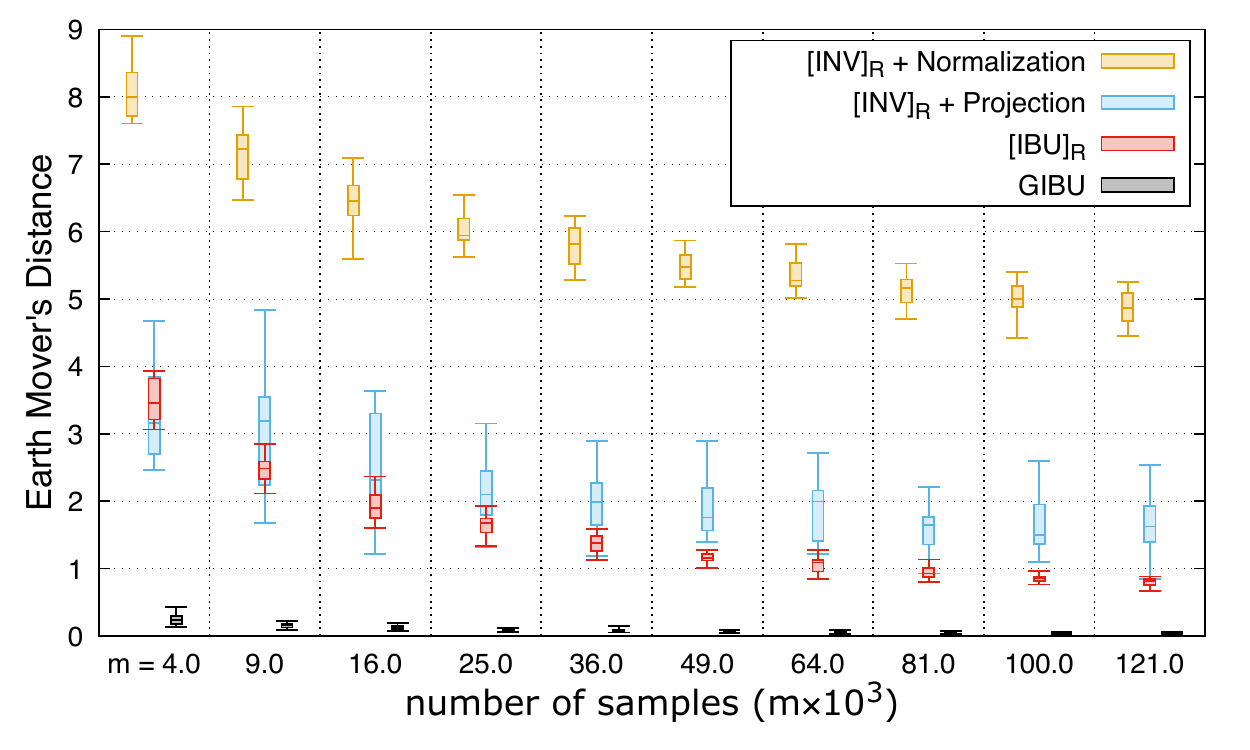}
\caption{Performance of \crinv{}, \cribu{} and \gibu{} for a linear alphabet of secrets $\{0,1,\dots,99 \}$
The noisy data are produced by $5$ truncated geometric mechanisms having $\geps_i$ and $5$ $k$-RR mechanisms (cfr.~\autoref{tab:geom_krr}). }
\label{fig:cr_VLGEOM_KRR}
\end{figure}

\begin{figure}[h]
\centering 
\includegraphics[width=0.50\textwidth]{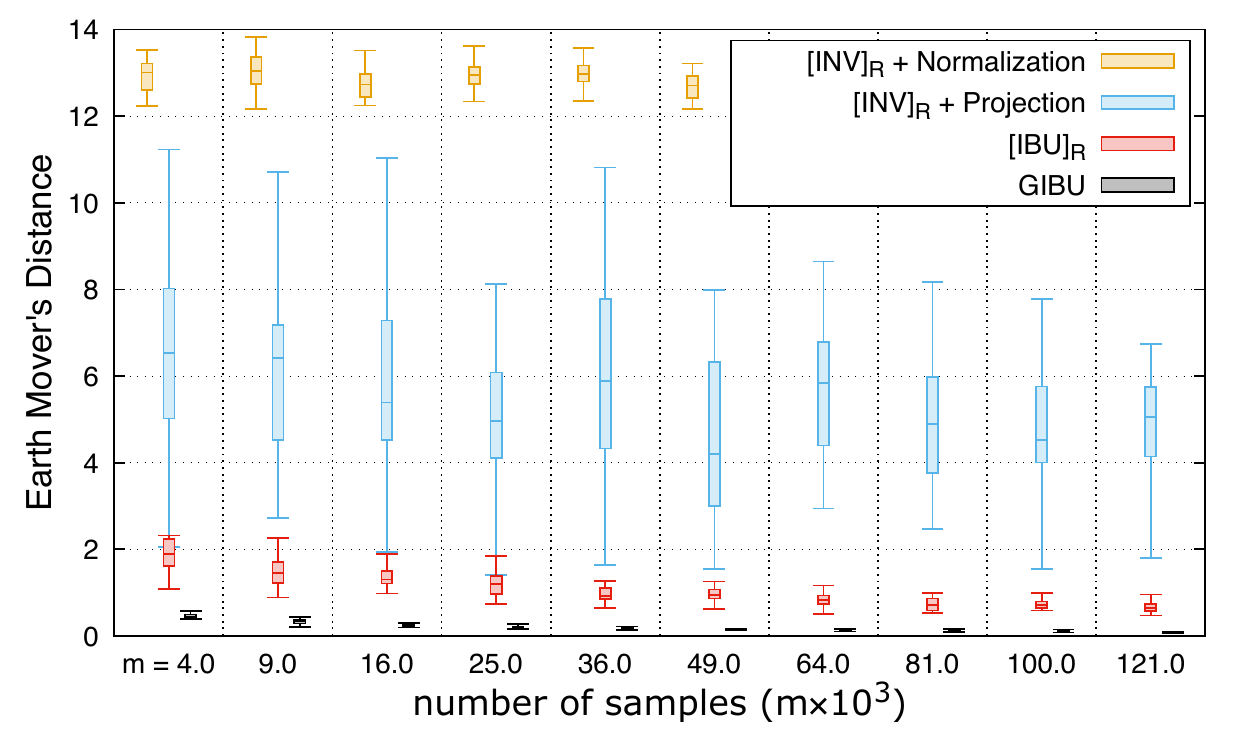}
\caption{Performance of \crinv{}, \cribu{} and \gibu{} for a linear alphabet of secrets $\{0,1,\dots,99 \}$
The noisy data are produced using 10 different linear geometric mechanisms (cfr.~\autoref{tab:geom_geom}).}
\label{fig:cr_GEOM_GEOM}
\end{figure}

\begin{figure}[h]
\centering 
\includegraphics[width=0.50\textwidth]{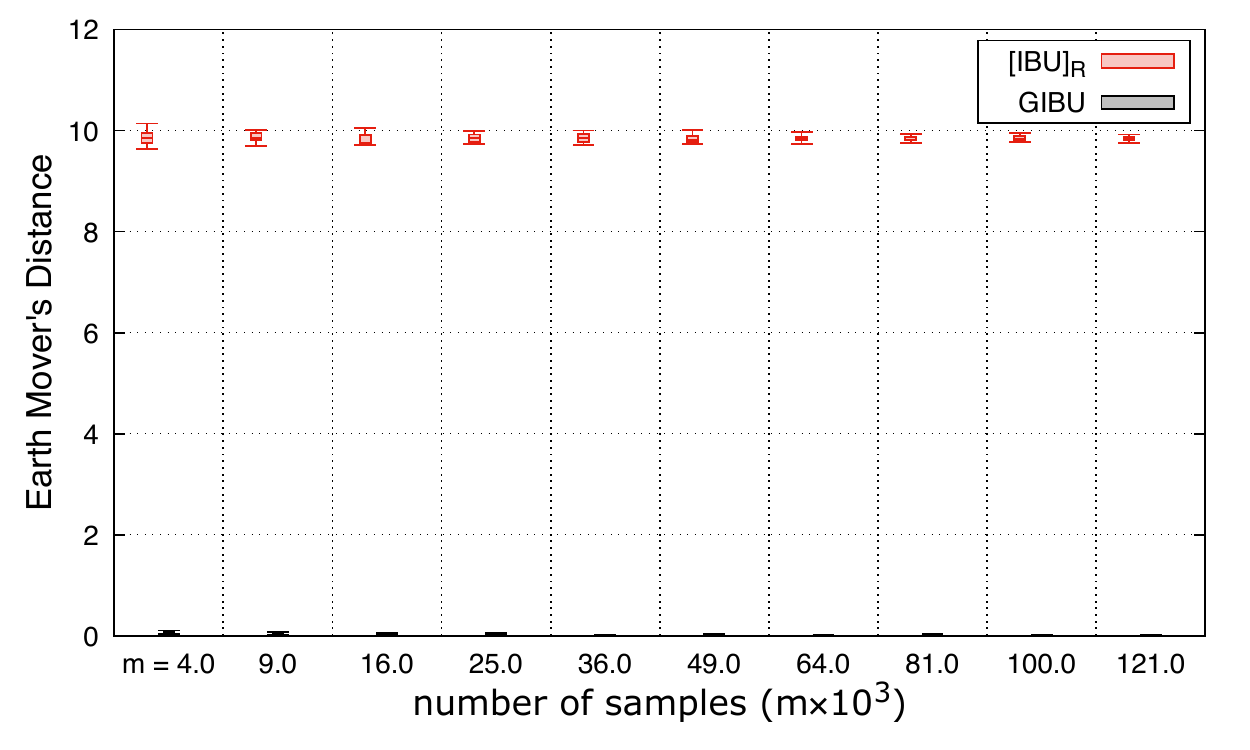}
\caption{Performance of \cribu{} and \gibu{} for a linear alphabet of secrets $\{0,1,\dots,99 \}$
The noisy data are produced using 10 different Shokri's mechanisms (cfr.~\autoref{tab:sh_sh}).}
\label{fig:cr_SH_SH}
\end{figure}

We also let the original data be obfuscated by $10$ $k$-RR mechanisms having $\leps_i$ 
varying between $3.0$ and $8.08$, and apply Equations (\ref{eq:generic_e}) and (\ref{eq:combining_results}) 
with $\mathbf{e}$ replaced by \inv{} and \ibu{}. We call the methods resulting from these two substitutions of $\mathbf{e}$ as 
\crinv{} and \cribu{} respectively. Similar to \rappor{}, every run of \inv{} requires post-processing by projection or normalization
to return a valid distribution (cfr.~Section \ref{sec:cminv}). 
Figure \ref{fig:cr_KRR_KRR} shows the performances of \crinv{} 
(with post-processing), \cribu{}, and \gibu{}.
We perform a similar experiment, but using a mixture of $5$ truncated geometric mechanisms with $\geps_i$ between $0.065$ and $0.869$  
and $5$ $k$-RR mechanisms with $\leps_i$ between $3.0$ and $4.69$,  
and we show the results in Figure \ref{fig:cr_VLGEOM_KRR}. 
Figure~\ref{fig:cr_GEOM_GEOM} shows the experiment results in the case of using a mixture of linear geometric mechanisms. 
Finally Figure~\ref{fig:cr_SH_SH} show the results for a mixtures of Shokri's mechanisms which we describe later in Section~\ref{sec:shokri:exp}. 

It is clear from Figures~\ref{fig:cr_KRR_KRR},\ref{fig:cr_VLGEOM_KRR},\ref{fig:cr_GEOM_GEOM},and \ref{fig:cr_SH_SH} that 
the method of combining results shows a significantly poor estimation quality compared to \gibu. This is explained by the fact that every underlying estimator 
works only on a \emph{small} subset of data $D^A$ instead of the entire set. This implies that the estimation error of $\mathbf{e}(A, D^A)$, for every $A$, is relatively large which makes the overall error of the final $\hat\vtheta$ also large. 

\section{Evaluation of combining mechanisms}

\subsection{Performance of \cminv{} relative to \gibu{}}\label{sec:inv_gibu}
In this section we experimentally compare between the estimation performance of the compound-mechanism 
inversion \cminv{} and \gibu{}. We will run our experiments on two alphabets of the users' private data. 
The first alphabet is \emph{linear} which may represent e.g. ages, smoking rates, etc. 
The other alphabet is \emph{planar} which represent geographic locations. 

In the linear case, we define the alphabet of secrets $\calx$ to be $\{0,1,\dots, 99\}$, with distance $1.0$ between 
successive elements. 
\footnote{Specifying a metric on $\calx$ is necessary to evaluate the earth mover's distance between the original and estimated distributions on $\calx$.} 
In this case, we synthesize the original data of the users by sampling from $\calx$ 
according to a binomial distribution with $\alpha=0.5$. 

For the planar alphabet, we consider a geographic region in San-Francisco bounded by the latitudes $37.7228$, $37.7946$, 
and the longitudes $-122.5153$, $-122.3789$. We partition this region into a grid of $24\times16$ cells, 
where the size of each cell is $0.5$km, and we define the alphabet of locations to be the set of these cells. The original 
data of the users are obtained from the Gowalla dataset, where each user datum is the cell that encloses his checkin. 
In the above two cases we will consider different mixtures of mechanisms applied by the users. 

\subsubsection{Various $k$-RR mechanisms}\label{sec:gibu_inv_krr}

\begin{figure}[h]
\centering 
\subfigure[linear space]{
      \label{fig:cm_KRR_L1.0_KRR_L28.0_INV_linear}
      \includegraphics[width=0.5\textwidth]{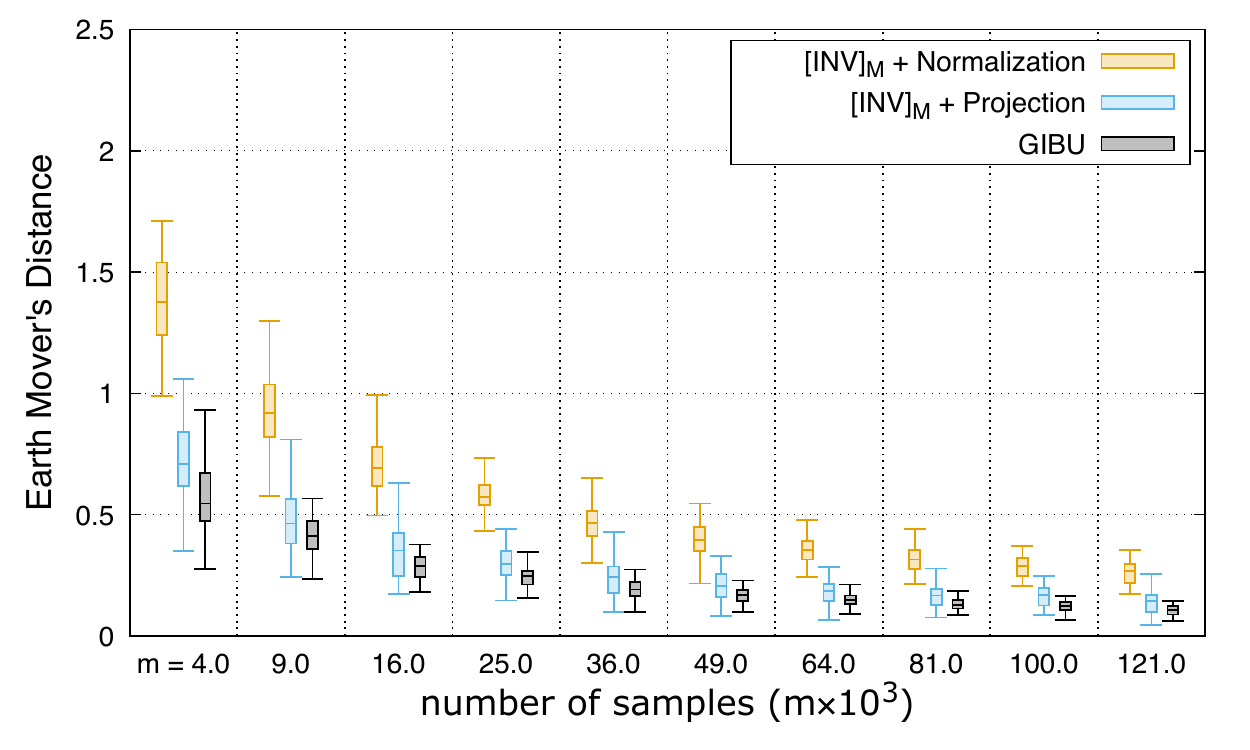}
      }
\subfigure[planar space]{
      \label{fig:cm_KRR_L0.5_KRR_L5.0_INV_planar}
      \includegraphics[width=0.5\textwidth]{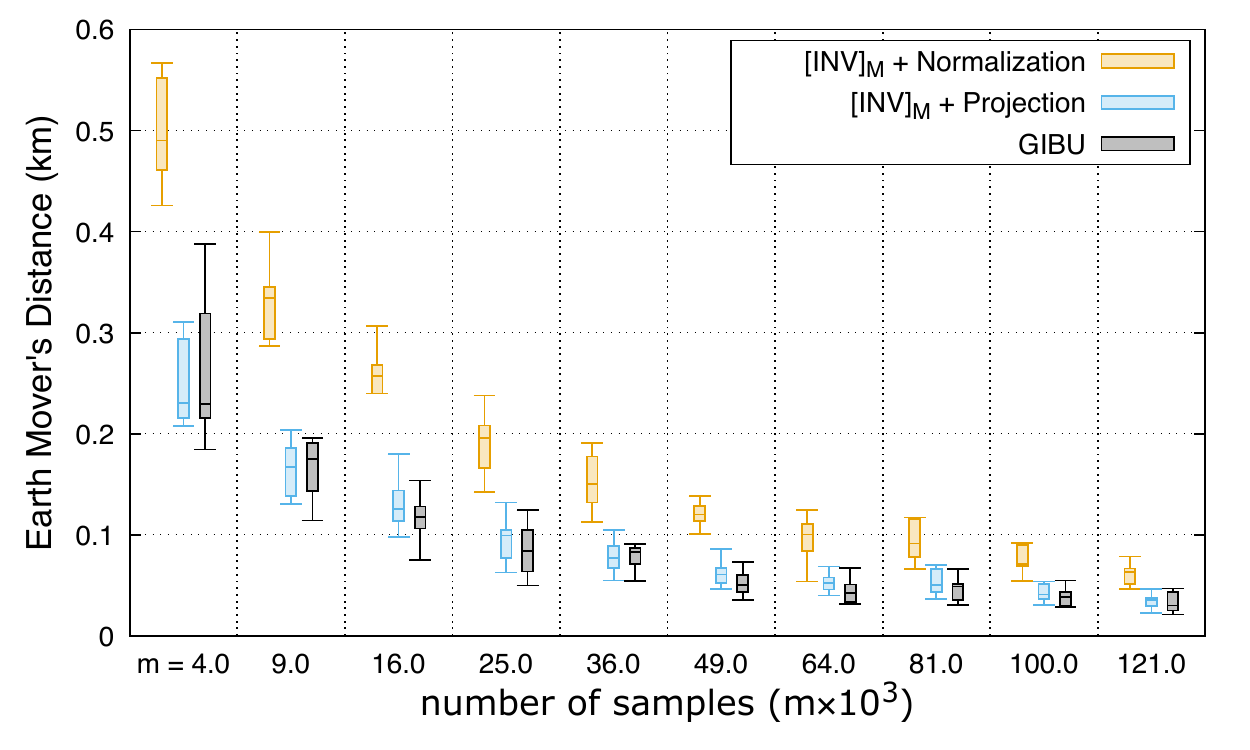}
      }
\caption{Performance of \cminv{} and \gibu{} under $10$ different $k$-RR mechanisms constructed for (a) the linear alphabet
and (b) the planar alphabet (cfr.~\autoref{tab:krr_krr}). }
\label{fig:cm_KRR_KRR_INV}
\end{figure}

We let the original data of users be obfuscated using $10$ $k$-RR mechanisms having various values of $\leps_i$. 
In the case of linear alphabet we set these values to be between $3.0$ and $8.08$. In the case of the planar 
alphabet we set the values of $\leps_i$ to be between $3.05$ and $8.20$. The complete list of these values 
are shown in \autoref{tab:krr_krr}.
Figure~\ref{fig:cm_KRR_KRR_INV} shows the estimation performance of \cminv{} 
with the two post-processing methods (normalization and projection), and also the performance of \gibu{}.  

From Figure \ref{fig:cm_KRR_KRR_INV} we observe first that for both linear and planar alphabets,  
post-processing the result of \cminv{} using projection exhibits significantly better estimation performance 
compared to normalization. We also observe that \gibu{} slightly outperforms \cminv{} with projection.

\subsubsection{Various geometric mechanisms}\label{sec:gibu_inv_geom}

\begin{figure}[h]
\centering 
\subfigure[linear alphabet]{
      \label{fig:cm_VLGEOM_L1.0_VLGEOM_L28.0_INV_linear}
      \includegraphics[width=0.5\textwidth]{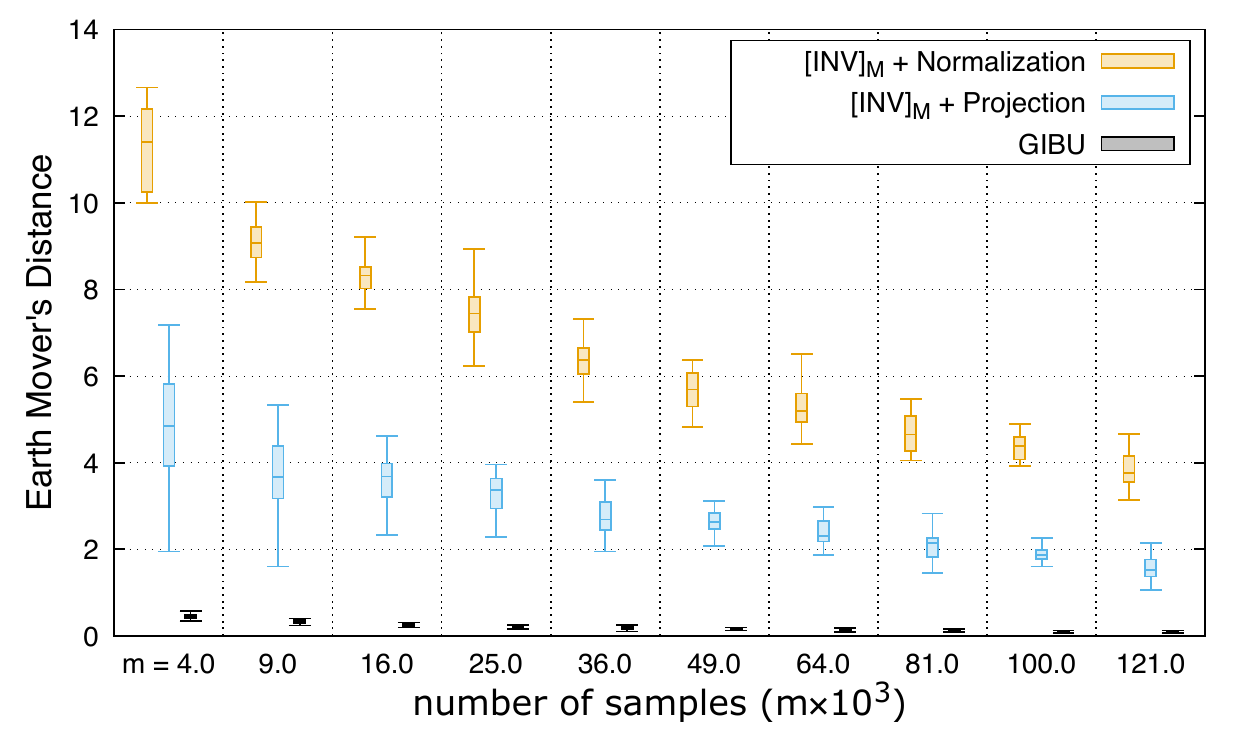}
      }
\subfigure[planar alphabet]{
      \label{fig:cm_PGEOM_L0.5_PGEOM_L5.0_INV_planar}
      \includegraphics[width=0.5\textwidth]{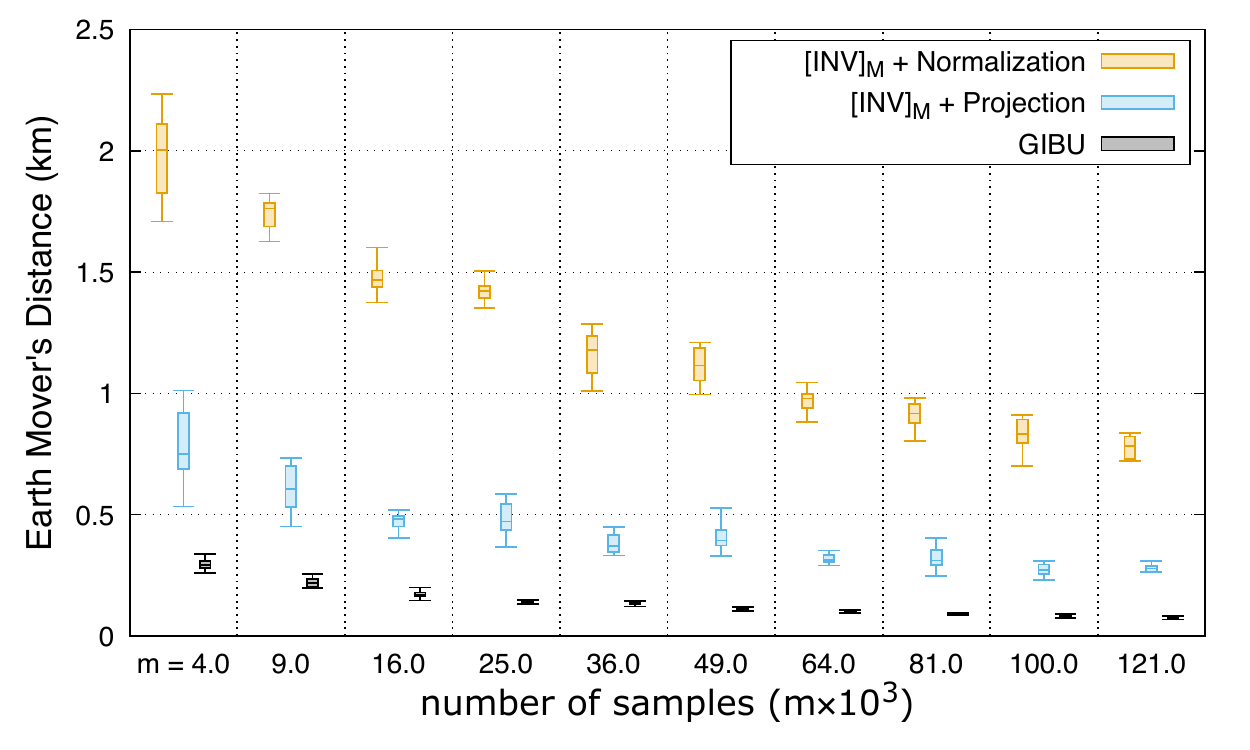}
      }
\caption{Performance of \cminv{} and \gibu{} under 10 geometric mechanisms constructed for 
(a) the linear alphabet and (b) the planar alphabet (cfr.~\autoref{tab:geom_geom}).}
\label{fig:cm_GEOM_GEOM_INV}
\end{figure}

We let the private data of the users be sanitized by the geometric mechanisms 
which satisfy $\geps$-geo-indistinguishability.  
For the linear alphabet we use 10 linear geometric mechanisms, defined by (\ref{eq:tgeometric}), with $\geps_i$ varying from  
$0.020$ to $0.869$. For the planar alphabet we use 10 \emph{planar} geometric mechanisms 
(described in Section \ref{sec:plan_geom}) with 
$\geps_i$ between $0.190$ and $3.124$. The complete list of these values are shown in \autoref{tab:geom_geom}. 
Based on the noisy data, the original distribution is estimated using 
\cminv{} and \gibu{} and we show their performances in Figure \ref{fig:cm_GEOM_GEOM_INV}. 

It is clear from Figure \ref{fig:cm_GEOM_GEOM_INV} that the performance of \gibu{} is significantly 
better than \cminv{} in both cases of linear and planar alphabets. 
%

\subsubsection{Mixed geometric and $k$-RR mechanisms}

\begin{figure}[h]
\centering 
\subfigure[linear alphabet]{
      \label{fig:cm_VLGEOM_L1.0_KRR_L28.0_INV_linear}
      \includegraphics[width=0.5\textwidth]{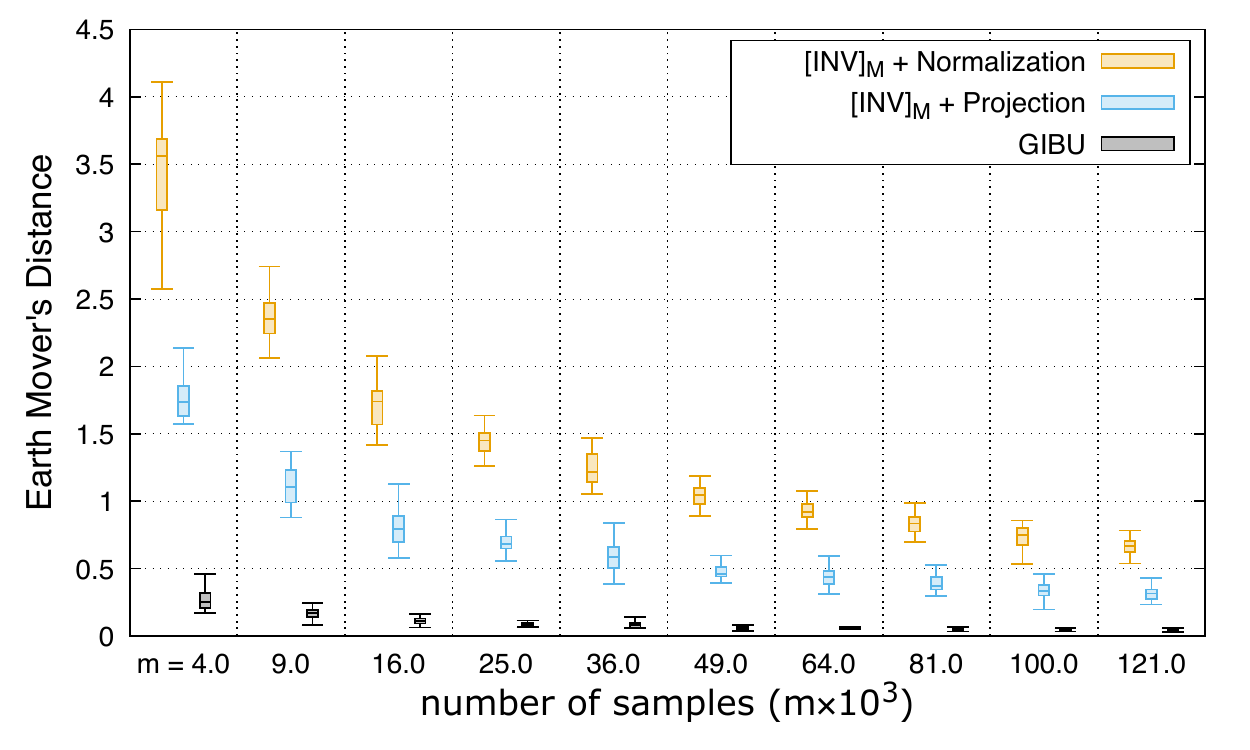}
      }
\subfigure[planar alphabet]{
      \label{fig:cm_PGEOM_L0.5_KRR_L5.0_INV_planar}
      \includegraphics[width=0.5\textwidth]{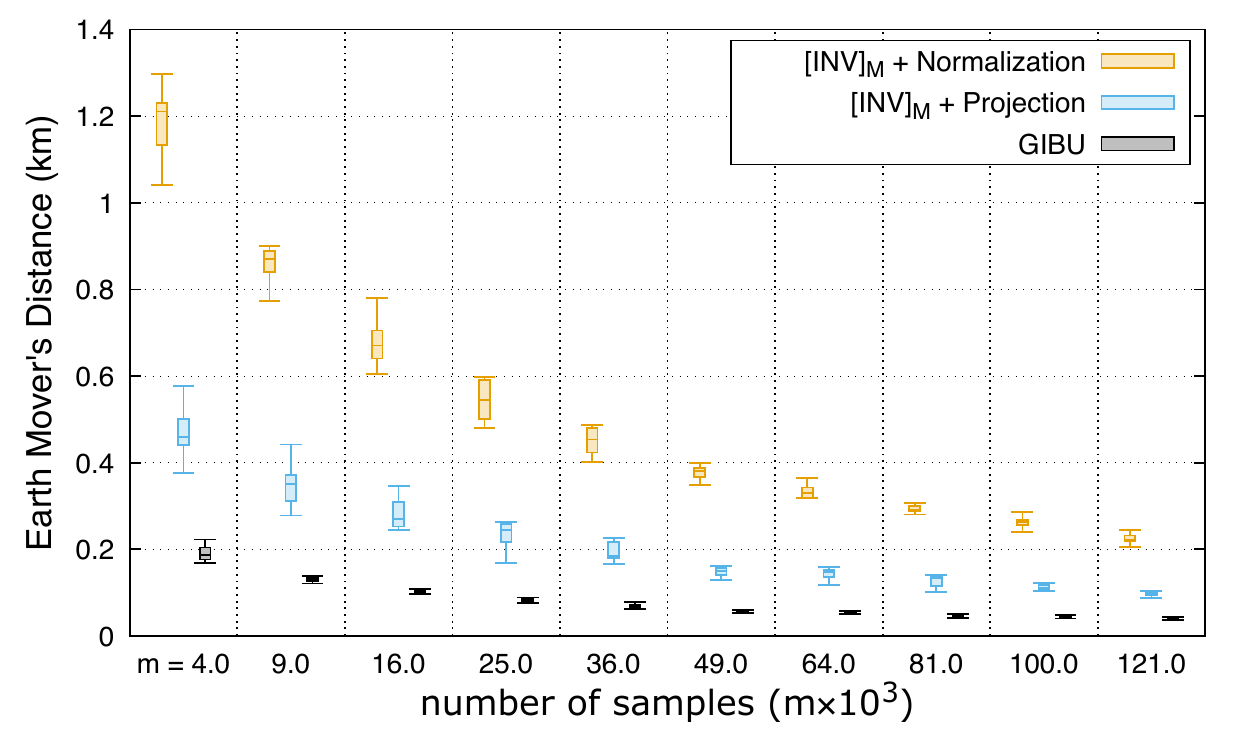}
      }
\caption{Performance of \cminv{} and \gibu{} under a mixture of $5$ different geometric mechanisms and $5$ different $k$-RR mechanisms
for (a) the linear alphabet, and (b) the planar alphabet (cfr.~\autoref{tab:geom_krr}). 
}
\label{fig:cm_GEOM_KRR_INV}
\end{figure}

We let now the original data of the users be sanitized by a mixture of geometric and $k$-RR mechanisms. 
Precisely, for the linear alphabet we let the original data be obfuscated by $5$ linear geometric mechanisms 
with $\geps_i$ varying between $0.065$ and $0.869$, and $5$ $k$-RR mechanisms with 
with $\leps_i$ varying between $3.0$ and $4.69$.
In the case of the planar alphabet we sanitize the original data by $5$ planar geometric mechanisms 
with $\geps_i$ between $0.632$ and $3.124$ and 5 $k$-RR mechanisms with $\leps_i$ 
between $3.05$ and $5.67$. The full list of these parameters is shown in \autoref{tab:geom_krr}. 
The results for \cminv{} and \gibu{}, in the two cases, are shown by Figure \ref{fig:cm_GEOM_KRR_INV}
which reflects again that \gibu{} provides a better estimation performance compared to \cminv{}. 
\subsection{Performance of \cmibu{} relative to \gibu{}}\label{sec:ibu_gibu}
In this section we compare between the performance of \cmibu{} described in Section \ref{sec:cmibu} and that of \gibu{}. 
We will use the same experimental setup of Section \ref{sec:inv_gibu}. 
In particular, we base our comparison on private data drawn from 
linear and planar alphabets. In the linear case, the alphabet is $\{0,1,\dots, 99\}$ and the private data of the users are synthesized using a binomial distribution. In the planar case, the alphabet is the grid cells of San Francisco and 
the private data are obtained from the Gowalla dataset. 


\subsubsection{Various $k$-RR mechanisms}\label{sec:gibu_ibu_krr}

\begin{figure}[h]
\centering 
\subfigure[linear space]{
	\label{fig:cm_KRR_L1.0_KRR_L28.0_IBU_linear}
	\includegraphics[width=0.5\textwidth]{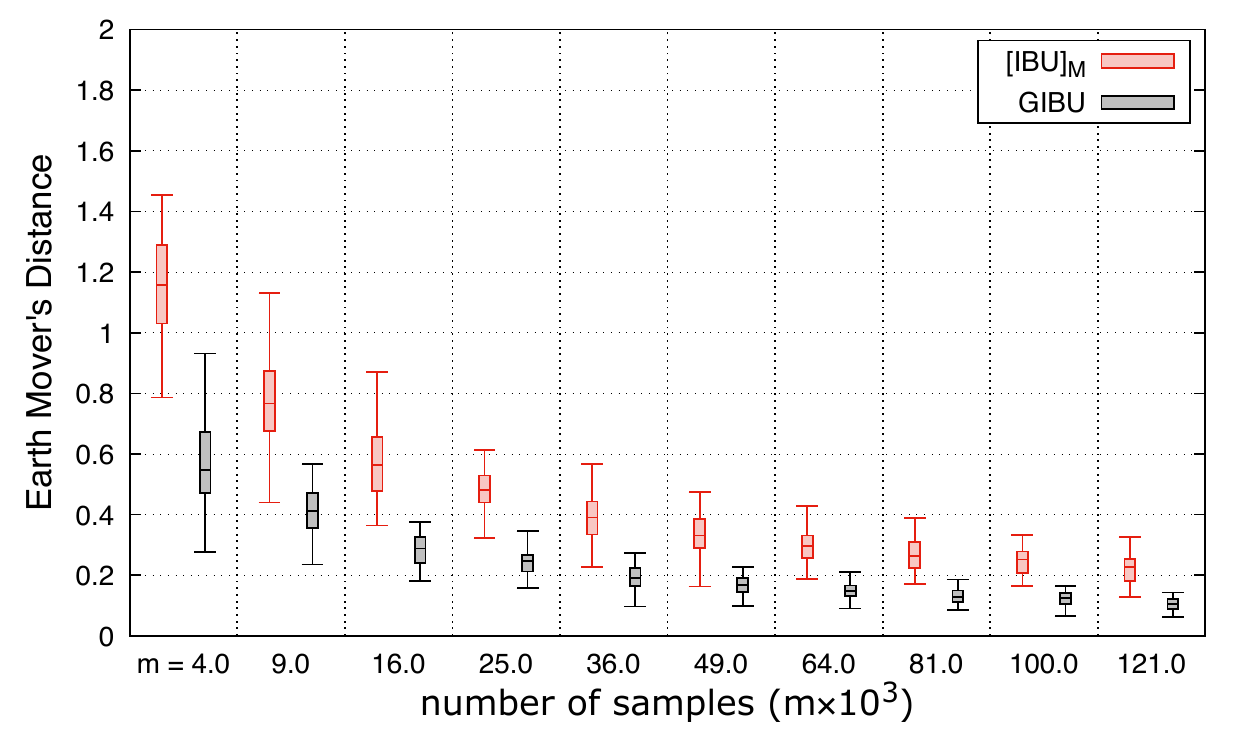}
	}
\subfigure[planar space]{
	\label{fig:cm_KRR_L0.5_KRR_L5.0_IBU_planar}
	\includegraphics[width=0.5\textwidth]{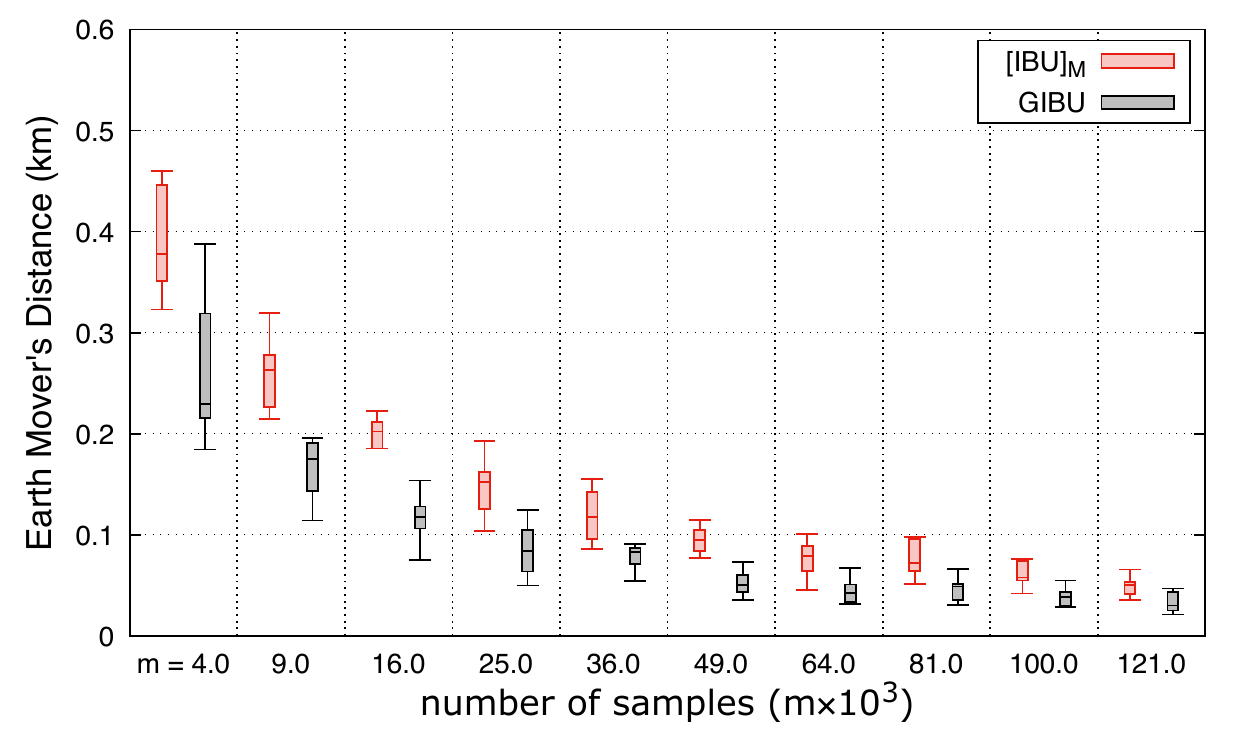}
	}
\caption{
Performance of \cmibu{} and \gibu{} under $10$ different $k$-RR mechanisms constructed for (a) the linear alphabet
and (b) the planar alphabet (cfr.~\autoref{tab:krr_krr}).
} 
\label{fig:cm_KRR_KRR_IBU}
\end{figure}

We let the original data of users be sanitized by 10 $k$-RR mechanisms with various values of $\leps_i$. 
In the case of linear alphabet, the values of $\leps_i$ are between $3.0$ to $8.08$, and in the case of planar 
alphabet, they are between $3.05$ and $8.20$ (cfr.~\autoref{tab:krr_krr}). 
In these two cases we estimate the original distribution on the alphabet 
using \cmibu{} and \gibu{} and show the results in Figure \ref{fig:cm_KRR_KRR_IBU}. 

It is clear that \gibu{} outperforms \cmibu{} in both cases of linear and
planar data. Note that although each one of these methods maximizes 
a likelihood function based on reported data, only \gibu{} yields the true MLE since it takes 
into account the various marginal distributions induced by individual mechanisms. \cmibu{} on the other hand 
returns the distribution under a `fake' 
hypothesis that all observations are 
produced by the average mechanism $A[n]$, hence returning a worse estimate. 

This also explains why the performance of \cmibu{} is worse than the performance of \cminv{} under 
$k$-RR mechanisms (cfr.~Figure \ref{fig:cm_KRR_KRR_INV}) although both methods use 
the average mechanism. In fact, unlike the case of \cmibu{}, the definition of \cminv{} is independent of 
the above hypothesis that observations are produced by $A[n]$. 

\subsubsection{Various geometric mechanisms}\label{sec:gibu_ibu_geom}

\begin{figure}[h]
\centering 
\subfigure[linear space]{
	\label{fig:cm_VLGEOM_L1.0_VLGEOM_L28.0_IBU_linear}
	\includegraphics[width=0.5\textwidth]{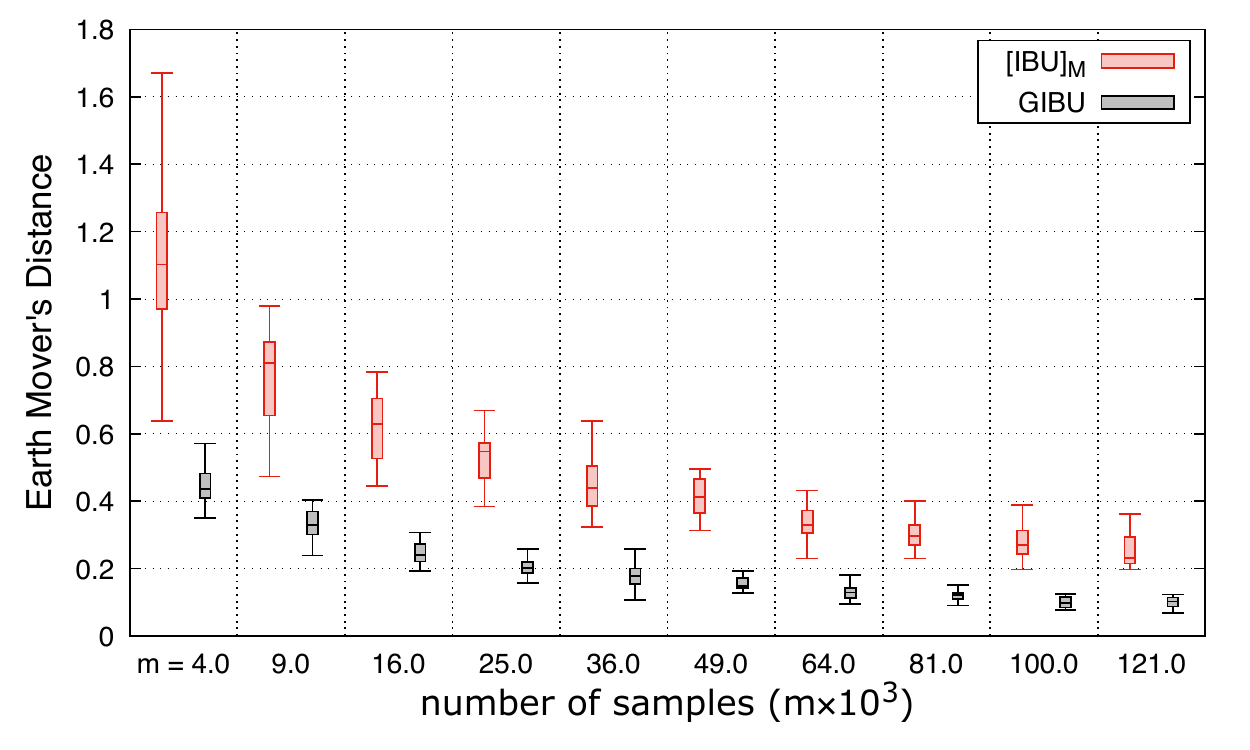}
	}
\subfigure[planar space]{
	\label{fig:cm_PGEOM_L0.5_PGEOM_L5.0_IBU_planar} 
	\includegraphics[width=0.5\textwidth]{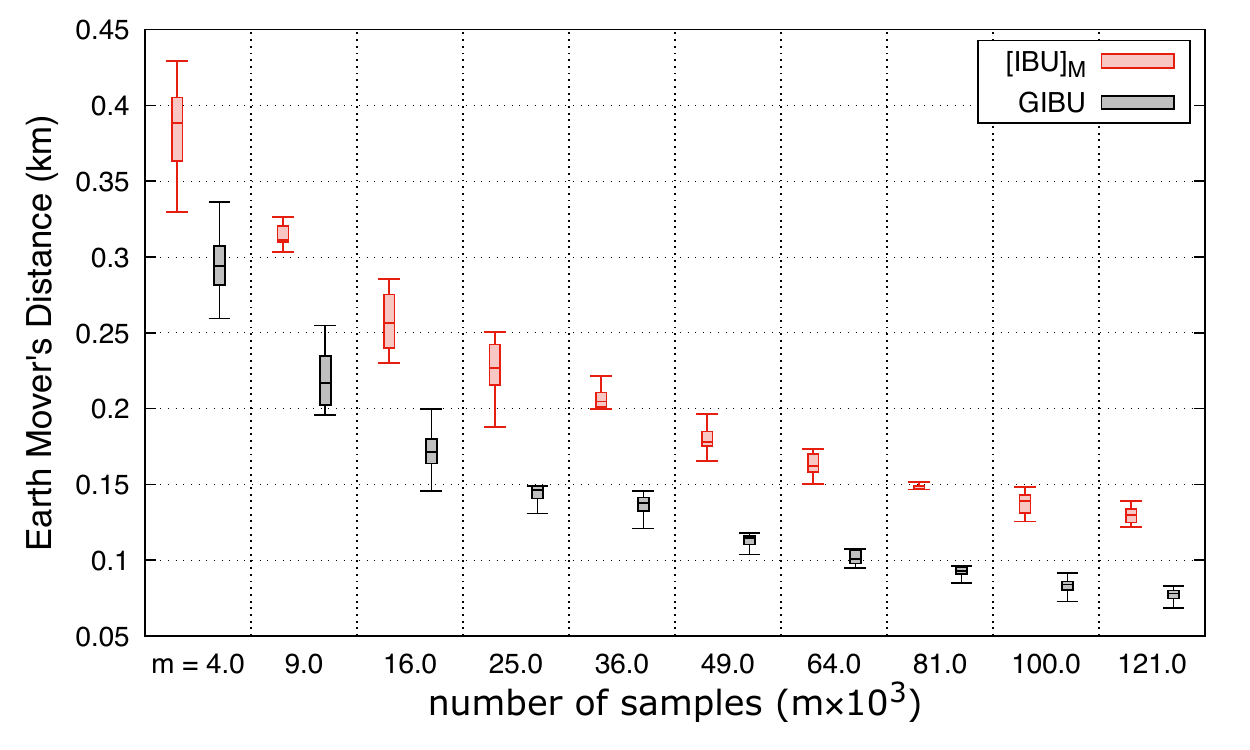}
	}
\caption{Performance of \cmibu{} and \gibu{} under $10$ different geometric mechanisms constructed for (a) the linear alphabet, 
and (b) the planar alphabet (cfr.~\autoref{tab:geom_geom}).}   
\label{fig:cm_GEOM_GEOM_IBU}
\end{figure}

Now we let the original data be sanitized by 10 different geometric geometric mechanisms.
For the linear alphabet we use linear geometric mechanisms with $\geps_i$ between $0.02$ and $0.869$, 
whereas for the planar alphabet we use the planar variants of geometric mechanisms with $\geps_i$ between 
$0.19$ and $3.124$ (cfr.~\autoref{tab:geom_geom}). 
Figure \ref{fig:cm_GEOM_GEOM_IBU} shows the estimation performances of \cmibu{} and \gibu{} in these
two cases. It is clear from this figure that \gibu{} is superior.

\subsubsection{Mixed geometric and $k$-RR mechanisms}

\begin{figure}[h]
\centering
\subfigure[linear space]{
	\label{fig:cm_VLGEOM_L1.0_KRR_L28.0_IBU_linear} 
	\includegraphics[width=0.5\textwidth]{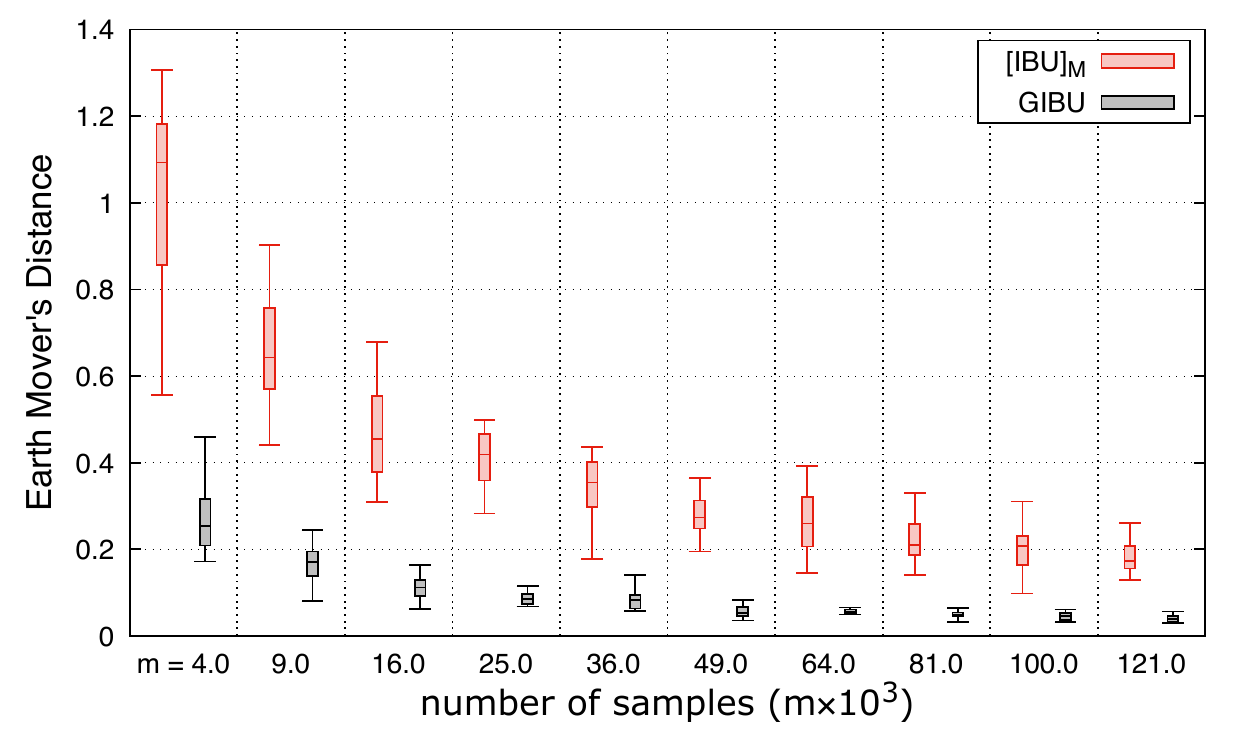}
	}
 \subfigure[planar space]{
 	\label{fig:cm_PGEOM_L0.5_KRR_L5.0_IBU_planar}
	\includegraphics[width=0.5\textwidth]{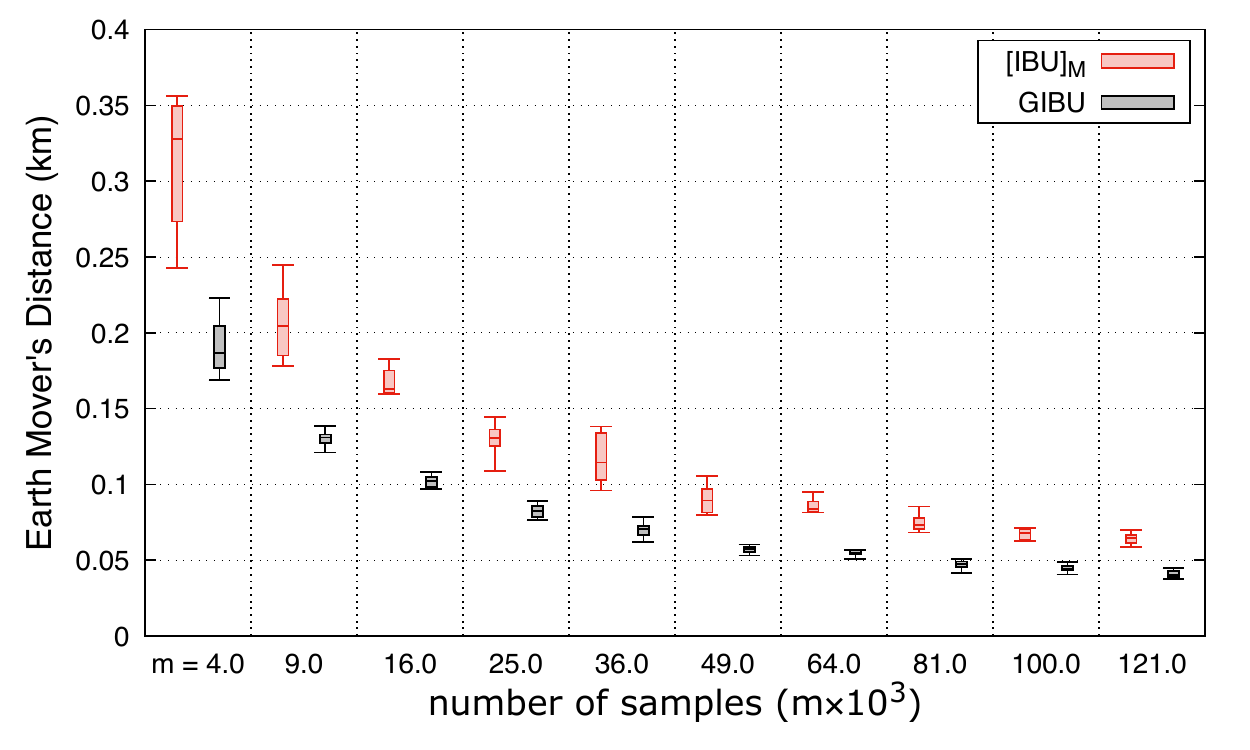}
	}
\caption{Performance of \cmibu{} and \gibu{} under a mixture of $5$ different geometric mechanisms and $5$ different $k$-RR mechanisms
for (a) the linear alphabet, and (b) the planar alphabet (cfr.~\autoref{tab:geom_krr}).} 
\label{fig:cm_GEOM_KRR_IBU}
\end{figure}

We consider now the scenario when both geometric and $k$-RR mechanisms are used by different users to obfuscate 
their data. In the case of the linear alphabet, we use $5$ linear geometric mechanism with 
various $\geps_i$ between $0.065$ and $0.869$ together with $5$ $k$-RR mechanisms with $\leps_i$ between $3.0$ and $4.69$. 
In the planar case we use $5$ planar geometric mechanisms with $\geps_i$ between $0.632$ and $3.124$ with $5$ $k$-RR 
mechanisms having $\leps_i$ between $3.05$ and $5.67$ (cfr.~\autoref{tab:geom_krr}). 
The performances of \cmibu{} and \gibu{} are shown by Figure \ref{fig:cm_GEOM_KRR_IBU} from which it is clear that \gibu{} is superior.  

\subsubsection{Privacy mechanisms of Shokri et al. }\label{sec:shokri:exp}

In the following we use the privacy mechanism of Shokri et al. \cite{Shokri:12:CCS} to obfuscate 
the original data of the users and again we compare the estimation performances of \cmibu{} and \gibu{}. 

This mechanism is defined 
so that the privacy, quantified  by the adversary's expected loss, is maximized while the expected loss 
of quality, experienced by the user, is maintained under a given threshold $Q$.
Suppose that the user's real datum is $x\in\calx$. Then for any $z\in\calx$ the function $\ell(x, z)$ quantifies 
the user's loss of quality when the mechanism reports $z$ instead of $x$. We let $\ell(x, z)$ describe also 
the adversary's loss when his guess is $z$. The user is assumed to have a personal profile modeled as a 
distribution $\vpi$ over the alphabet $\calx$. The adversary is assumed to know $\vpi$ which he 
uses, in addition to the mechanism $A$, to make his guess so that his expected loss is minimized. 
It is shown in \cite{Shokri:12:CCS} that the adversary's best guess, given an observation $y$, is 
\[
\argmin_{z\in\calx} \sum_{x\in\calx} \pi_x A_{x y} \ell(x, z).
\] 
Hince the mechanism $A$ is defined to be the solution of the following optimization problem. 
\begin{align*}
\text{Maximize} \quad &\sum_{y\in\calx} \min_{z\in\calx} \sum_{x\in\calx} \pi_x A_{x y} \, \ell(x, z), \\
\text{subject to} \quad &\sum_{x\in\calx} \pi_x \sum_{z\in\calx} A_{xz} \ell(x,z) \leq Q, \\
&\sum_{z\in\calx} A_{x z} =1 \quad\forall x\in\calx, \\
& A_{x z} \geq 0 \quad \forall x, z \in \calx. 
\end{align*}
Note that the objective function above is the expected adversary's loss using his best strategy, 
and the first constraint restricts the user's expected loss of quality to be below the threshold $Q$. 

\begin{figure}[h]
\centering 
\subfigure[linear space]{
      \label{fig:SH1.0_SH28.0_IBU_linear}
      \includegraphics[width=0.5\textwidth]{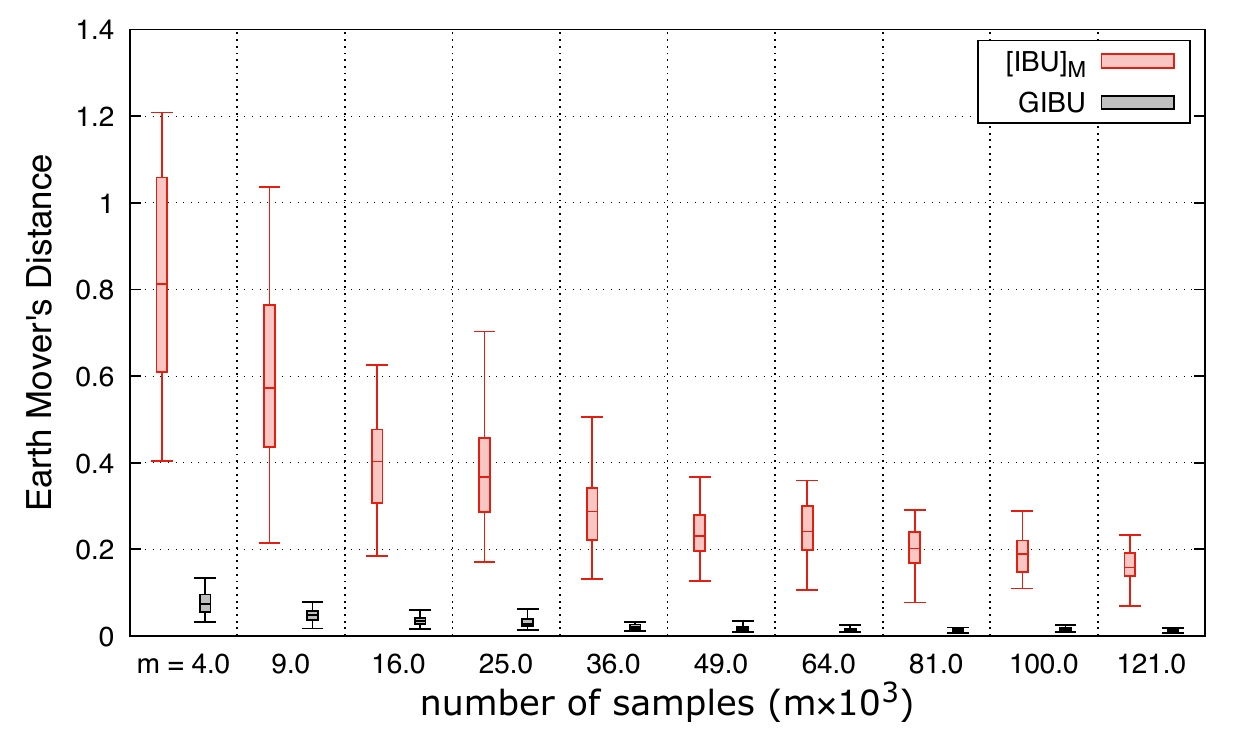}
      }
\subfigure[planar space]{
      \label{fig:SH0.5_SH14.0_IBU_planar}
      \includegraphics[width=0.5\textwidth]{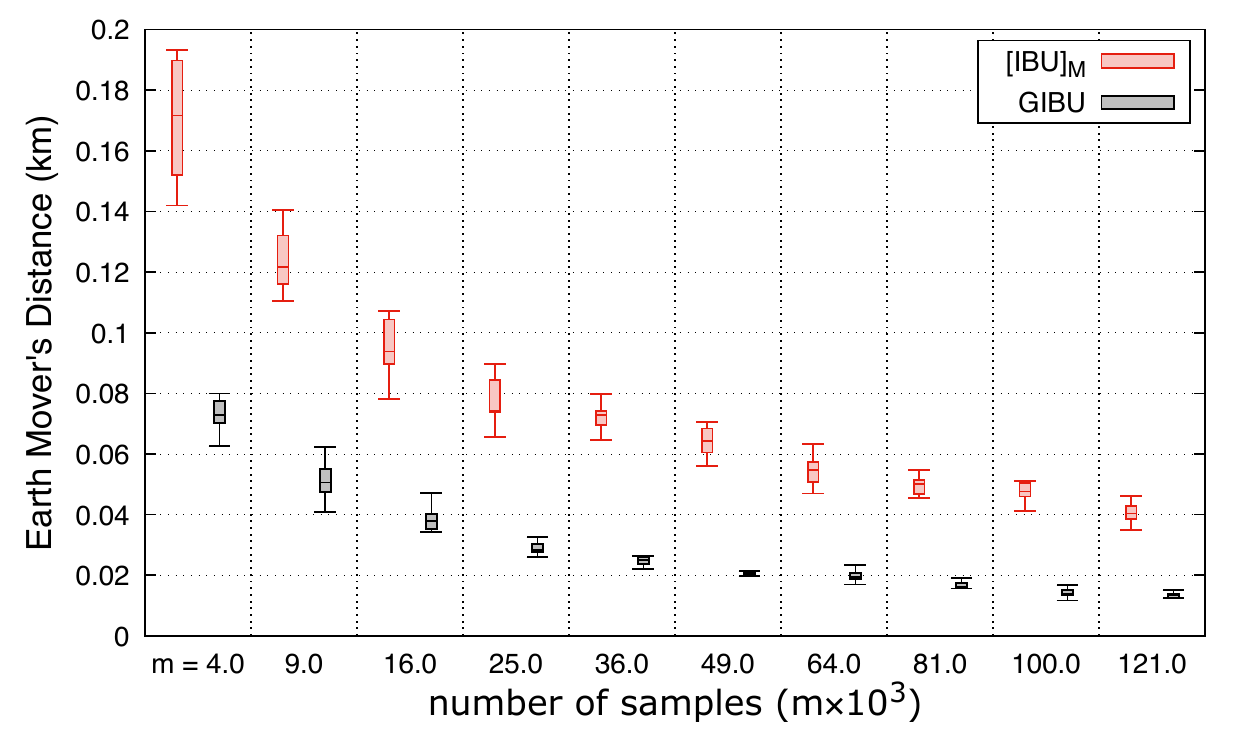}
      }
\caption{Performance of \cmibu{} and \gibu{} under $10$ different mechanisms of Shokri et al. \cite{Shokri:12:CCS}
for (a) the linear alphabet, and (b) the planar alphabet. The values $Q_i$ are set as in \autoref{tab:sh_sh}, and 
the public profiles $\vpi$ of the users are uniform.}   
\label{fig:cm_SH_IBU}
\end{figure}

Now for our experiment we let the original data of the users be obfuscated by $10$ Shokri's mechanisms 
constructed with $\vpi$ set to the uniform distribution on $\calx$, the loss function $\ell(\cdot, \cdot)$ set 
to be the Euclidean distance between the elements of $\calx$, and finally with various $Q_i$ as in \autoref{tab:sh_sh}. 
The performances of \cmibu{} and \gibu{} for the linear and planar alphabets are shown in 
Figure \ref{fig:cm_SH_IBU} which reflects the superiority of \gibu{} in this experiment. 

\subsection{Performance of \cmrappor{} relative to \gibu{}}\label{sec:rappor_gibu}
We suppose that original private data take values from a linear alphabet $\calx = \{0,1, \dots, 19\}$ in which the 
distance between any two successive elements is $1$.
The original data are synthetically generated according to a binomial 
distribution on $\calx$ with $\alpha=0.5$. Then the data are obfuscated using $10$ \textsc{Rappor} mechanisms. 

\begin{figure}[h]
\centering 
\subfigure[high privacy regime: $\leps_i$ between $0.1$ and $1.0$]{
      \label{fig:cm_RAP_P0.1_RAP_P1.0}
      \includegraphics[width=0.5\textwidth]{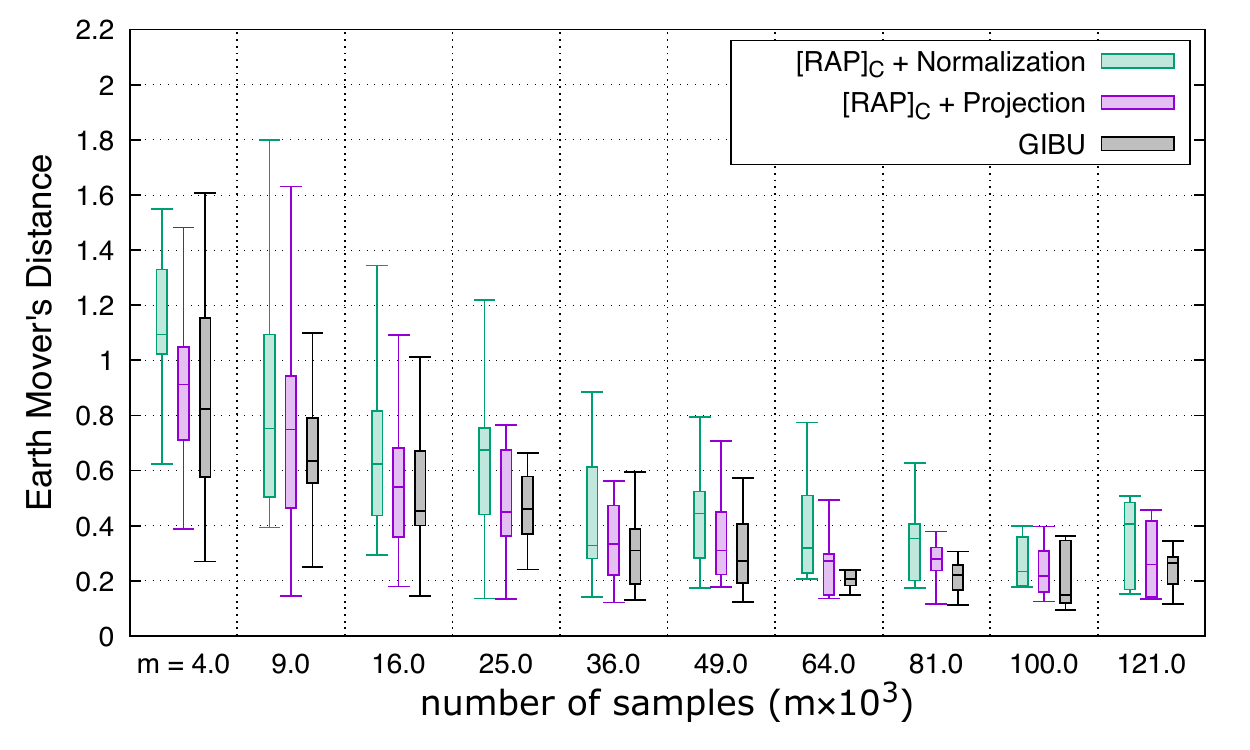}
      }
\subfigure[low privacy regime: $\leps_i$ between $1.0$ and $10.0$]{
      \label{fig:cm_RAP_P1.0_RAP_P10.0}
      \includegraphics[width=0.5\textwidth]{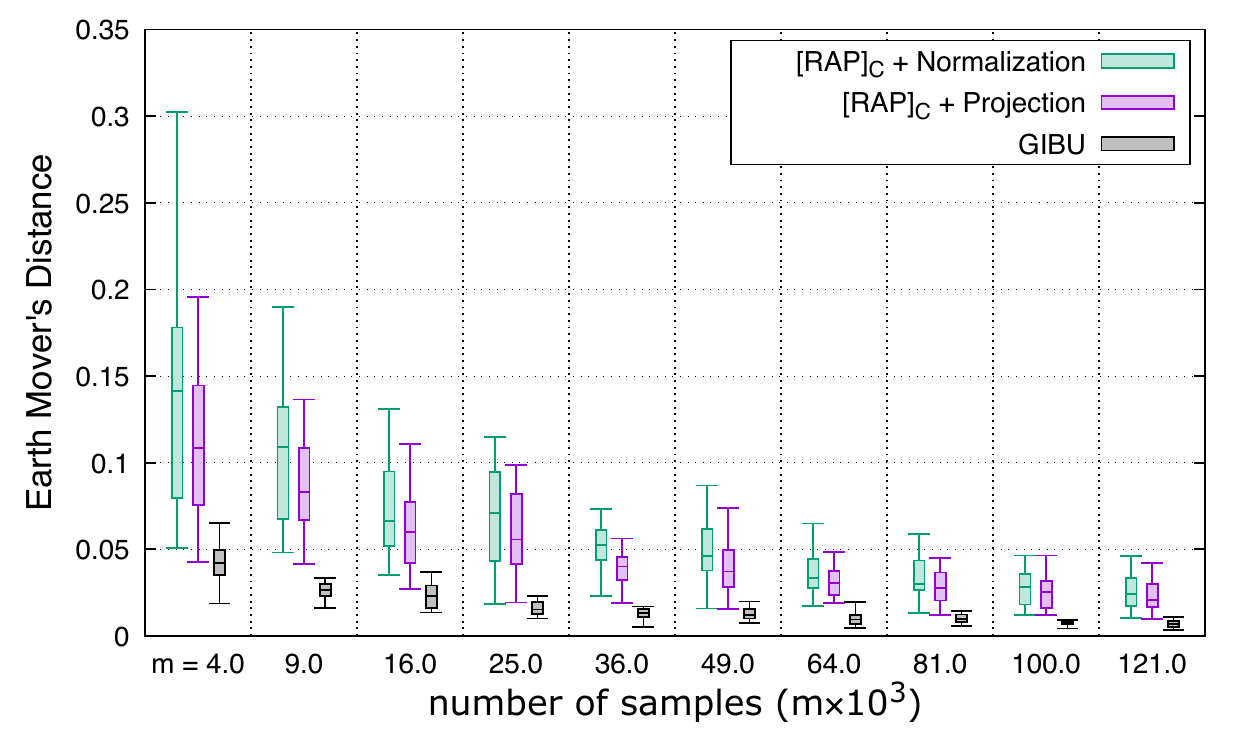}
      }
\caption{Performance of \cmrappor{} and \gibu{} for the alphabet of secrets $\{0,1,\dots,19 \}$.
The noisy data are produced by 10 \textsc{Rappor} mechanisms having various $\leps_i$ 
in high privacy (a), and in low privacy (b) (\emph{c.f} \autoref{tab:rappor_rappor}).}
\label{fig:cm_RAPPOR_RAPPOR}
\end{figure}

We perform our evaluation in two settings. In the first one, which we refer to as the \emph{high privacy} regime, 
we set $\leps_i$ of the mechanisms to vary between $0.1$ and $1.0$. In the 
other setting, which we call \emph{low privacy} regime, we set $\leps_i$ to vary between $1.0$ and $10.0$.  
In these two settings, Figure~\ref{fig:cm_RAPPOR_RAPPOR} shows the performance of \cmrappor{} with the two post-processing methods (normalization and projection), and also the performance of \gibu{}.

It is clear that the performances of \cmrappor{} and \gibu{} are similar in the high privacy regime, while the latter is 
clearly better in the low privacy regime. A related result obtained by \cite{Kairouz:16:ICML} states that
\textsc{Rappor} estimator (which assumes that all users apply the same mechanism) is `order' optimal when 
$\leps < 1$; while being not optimal for larger values of $\leps$. Our evaluation extends this result to 
\cmrappor{} which accepts various mechanisms. 
\section{Conclusion}

In this paper we consider the situation when every user applies his own privacy mechanism and privacy level to sanitize his sensitive data. Of course this situation involves many mechanisms that may differ in the level of privacy, in their signature, or in both. In this case we require to construct the probability distribution of the users' original data on the alphabet of the secrets. 
We have presented several methods for this construction and presented experimental comparisons between them. These comparisons were based on both synthetic and real datasets. We find that methods that are based on composing the mechanisms are more accurate than the methods based on combining results. Furthermore, we find that \gibu{} has a superior accuracy compared to other methods.


\bibliographystyle{ieeetran}
\bibliography{new_refs}

\appendices

\section{}\label{sec:mix_params}
\begin{table}[h!]
  \begin{center}
    \caption{\textsc{Rappor} Mechanisms used in Figures \ref{fig:cr_RAPPOR_RAPPOR} and \ref{fig:cm_RAPPOR_RAPPOR}}
    \label{tab:rappor_rappor}
    \bgroup
    \def\arraystretch{1.0}
    \begin{tabular}{ c | c c c c c }
    \toprule 
    \multirow{3}{*}{\bf{high privacy}} &\multicolumn{5}{c}{10 \textsc{Rappor} mechanisms with $\leps_i$ below} \\ 
    & 0.1 & 0.2 & 0.3 & 0.4 & 0.5 \\ 
    & 0.6 & 0.7 & 0.8 & 0.9 & 1.0 \\    
    \midrule 
    \multirow{3}{*}{\bf{low privacy}} &\multicolumn{5}{c}{10 \textsc{Rappor} mechanisms with $\leps_i$ below} \\ 
    & 1.0 & 2.0 & 3.0 & 4.0 & 5.0 \\ 
    & 6.0 & 7.0 & 8.0 & 9.0 & 10.0  \\    
    \bottomrule 
    \end{tabular}
    \egroup
  \end{center}
\end{table}

\begin{table}[h!]
  \begin{center}
    \caption{Mechanisms used in Figures \ref{fig:cr_KRR_KRR}, \ref{fig:cm_KRR_KRR_INV}, and \ref{fig:cm_KRR_KRR_IBU}} 
    \label{tab:krr_krr} 
    \bgroup 
    \def\arraystretch{1.0}   
    \begin{tabular}{ c | c c c c c  } 
    \toprule 
    \multirow{3}{*}{\bf{linear}} &  \multicolumn{5}{c}{10 $k$-RR mechanisms with $\leps_i$ below}\\ 
    & 3.00 & 3.54 & 3.96 & 4.34 & 4.69 \\  
    & 5.06 & 5.46 & 5.93 & 6.60 & 8.08 \\             
    \midrule 
    \multirow{3}{*}{\bf{planar}} &\multicolumn{5}{c}{10 $k$-RR mechanisms with $\leps_i$ below} \\ 
    & 3.05 & 4.19 & 4.81 & 5.27 & 5.67 \\
    & 6.05 & 6.44 & 6.87 & 7.40 & 8.20  \\ 
    \bottomrule 
    \end{tabular}
    \egroup
  \end{center}
\end{table}

\begin{table}[h!]
  \begin{center}
    \caption{Mechanisms used in Figures \ref{fig:cm_GEOM_GEOM_INV}, \ref{fig:cm_GEOM_GEOM_IBU}, and \ref{fig:cr_GEOM_GEOM}}
    \label{tab:geom_geom}
    \bgroup
    \def\arraystretch{1.0}
    \begin{tabular}{ c | c c c c c }
    \toprule 
    \multirow{3}{*}{\bf{linear}} &\multicolumn{5}{c}{10 truncated geometric mechanisms with $\geps_i$ below} \\ 
    & 0.020 & 0.025 & 0.031 & 0.039 & 0.050 \\
    & 0.065 & 0.088 & 0.131 & 0.236 & 0.869 \\    
    \midrule 
    \multirow{3}{*}{\bf{planar}} &\multicolumn{5}{c}{10 truncated planar mechanisms with $\geps_i$ below} \\ 
    & 0.190 & 0.244 & 0.310 & 0.390 & 0.493 \\
    & 0.632 & 0.835 & 1.159 & 1.762 & 3.124  \\    
    \bottomrule 
    \end{tabular}
    \egroup
  \end{center}
\end{table}

\begin{table}[h!]
  \begin{center}
    \caption{Mechanisms used in Figures \ref{fig:cr_VLGEOM_KRR}, \ref{fig:cm_GEOM_KRR_INV} and \ref{fig:cm_GEOM_KRR_IBU}}
    \label{tab:geom_krr}
    \bgroup
    \def\arraystretch{1.0}
    \begin{tabular}{ c | c c c c c }
    \toprule 
    \multirow{4}{*}{\bf{linear}} &\multicolumn{5}{c}{5 truncated geometric mechanisms with $\geps_i$ below} \\ 
    & 0.065 & 0.088 & 0.131 & 0.236 & 0.869 \\
    \cmidrule{2-6} 
    &\multicolumn{5}{c}{5 $k$-RR mechanisms with $\leps_i$ below} \\ 
    & 3.00 & 3.54 & 3.96 & 4.34 & 4.69 \\ 
    \midrule 
    \multirow{4}{*}{\bf{planar}} &\multicolumn{5}{c}{5 truncated planar mechanisms with $\geps_i$ below} \\ 
    & 0.632 & 0.835 & 1.159 & 1.762 & 3.124  \\    
    \cmidrule{2-6} 
    &\multicolumn{5}{c}{5 $k$-RR mechanisms with $\leps_i$ below} \\ 
    & 3.05 & 4.19 & 4.81 & 5.27 & 5.67 \\
    \bottomrule 
    \end{tabular}
    \egroup
  \end{center}
\end{table}

\begin{table}[h!]
  \begin{center}
    \caption{Mechanisms used in Figures \ref{fig:cm_SH_IBU}}
    \label{tab:sh_sh}
    \bgroup
    \def\arraystretch{1.0}
    \begin{tabular}{ c | c c c c c }
    \toprule 
    \multirow{3}{*}{\bf{linear}} &\multicolumn{5}{c}{10 Shokri's mechanisms with $Q_i$ below} \\ 
    & 1.0 & 4.0 & 7.0 & 10.0 & 13.0 \\
    &  16.0 & 19.0 & 22.0 & 24.5 & 28.0  \\    
    \midrule 
    \multirow{3}{*}{\bf{planar}} &\multicolumn{5}{c}{10 Shokri's mechanisms with $Q_i$ below} \\ 
    & 0.3 & 0.6 & 0.9 & 1.2 & 1.5 \\
    & 1.8 & 2.1 & 2.4 & 2.7 & 3.0 \\    
    \bottomrule 
    \end{tabular}
    \egroup
  \end{center}
\end{table}

\end{document}